\documentclass[conference]{IEEEtran}

\usepackage{tikz}
\usepackage{amsmath}
\usepackage{filecontents}
\usepackage{booktabs}
\usepackage{multicol} 
\usepackage{array}
\usepackage{arydshln}
\usepackage{subfig,graphicx}
\usepackage{multirow}
\usepackage{setspace}
\usepackage{array}
\usepackage{makecell}
\usepackage{balance}
\usepackage{algpseudocode}
\usepackage[linesnumbered]{algorithm2e}
\usepackage{soul} 
\usepackage{color, xcolor} 
\usepackage{lettrine}
\usepackage{picinpar}
\usepackage{sidecap}

\def\ie{\textit{i.e.}\xspace}
\def\etc{\textit{etc}\xspace}
\def\eg{\textit{e.g.}\xspace}
\def\etal{\textit{et~al.}\xspace}   
\def\red1#1{{\color{black}#1}}

\ifCLASSINFOpdf
 
\else
  
\fi

\begin{document}
%

\title{\Large \bf PrintListener: Uncovering the Vulnerability of Fingerprint Authentication via the Finger Friction Sound}

\author{\IEEEauthorblockN{
		 Man Zhou$^{1\dagger*}$, Shuao Su$^{1\dagger}$, Qian Wang$^{2}$, Qi Li$^{3}$, Yuting Zhou$^{1\dagger}$, Xiaojing Ma$^{1\dagger}$ and Zhengxiong Li$^{4}$}
	\IEEEauthorblockA{$^{1}$School of Cyber Science and Engineering, Huazhong University of Science and Technology, China}
	\IEEEauthorblockA{$^{2}$School of Cyber Science and Engineering, Wuhan University, China}	\IEEEauthorblockA{$^{3}$Institute for Network Sciences and Cyberspace, Tsinghua University, China}	\IEEEauthorblockA{$^{4}$Department of Computer Science and Engineering, University of Colorado Denver, USA}
	Email: \{zhouman, m202271728, m202271743, lindahust\}@hust.edu.cn, \\
	qianwang@whu.edu.cn, qli01@tsinghua.edu.cn, zhengxiong.li@ucdenver.edu	 
}

\IEEEoverridecommandlockouts
\makeatletter\def\@IEEEpubidpullup{6.5\baselineskip}\makeatother
\IEEEpubid{\parbox{\columnwidth}{
		Network and Distributed System Security (NDSS) Symposium 2024\\
		26 February - 1 March 2024, San Diego, CA, USA\\
		ISBN 1-891562-93-2\\
		https://dx.doi.org/10.14722/ndss.2024.24618\\
		www.ndss-symposium.org
	}
	\hspace{\columnsep}\makebox[\columnwidth]{}}

\maketitle

\renewcommand{\thefootnote}{\fnsymbol{footnote}}
\setcounter{footnote}{0}

\footnotetext[1]{Corresponding author: Man Zhou (zhouman@hust.edu.cn).}
\footnotetext[2]{Hubei Key Laboratory of Distributed System Security, Hubei Engineering Research Center on Big Data Security, School of Cyber Science and Engineering, Huazhong University of Science and Technology.}

\begin{abstract}
Fingerprint authentication has been extensively employed in contemporary identity verification systems owing to its rapidity and cost-effectiveness. Due to its widespread use, fingerprint leakage may cause sensitive information theft, enormous economic and personnel losses, and even \red1{a potential compromise of} national security. As a fingerprint that can coincidentally match a specific proportion of the overall fingerprint population, MasterPrint rings the alarm bells for the security of fingerprint authentication. 
In this paper, we propose a new side-channel attack on the minutiae-based Automatic Fingerprint Identification System (AFIS), called PrintListener, which leverages users' fingertip swiping actions on the screen to extract fingerprint pattern features (the first-level features) and synthesizes a stronger targeted PatternMasterPrint with potential second-level features. The attack scenario of PrintListener is extensive and covert. 
It only needs to record users' fingertip friction sound and can be launched by leveraging a large number of social media platforms. Extensive experimental results in real-world scenarios show that Printlistener can \red1{significantly} improve the attack potency of MasterPrint.

\end{abstract}

\section{Introduction}

The fingerprint is an impression left by the friction ridges of a human finger. 
As an essential and well-recognized personal identification (ID), the fingerprint has seamlessly integrated into people's daily life, \eg, phone screen unlock~\cite{Fingerrate}, fingerprint online payments~\cite{FingerPay}, national ID cards/electronic passports~\cite{passport}, \etc. 
Consequently, it has gained widespread adoption in applications such as user authentication, online transaction platforms, access control, and government and law enforcement agencies. 
It is projected that the market size of fingerprint authentication will touch USD 99.9 billion by 2032~\cite{Fingerprint}.




Preventing user fingerprint information leakage is a significant challenge worldwide ~\cite{zhang2015fingerprints}. 
The breach of such information can lead to the theft of sensitive data, substantial financial and human losses, and even pose a threat to national security. The current fingerprint attack explorations are mainly based on contact scrutinization. It is possible to extract usable fingerprints from the surface touched ~\cite{fisher2012techniques} or photos of exposed fingers ~\cite{echizen2018biometricjammer}.
Therefore, one intuitive suggestion to enhance fingerprint security is that people can keep their fingerprints away from others' sight, \eg, no adversary proximity/accessibility and no access to their devices.
However, is this ideal scenario truly secure against attacks? As a fingerprint that can coincidentally match a specific proportion of the overall fingerprint population, MasterPrint ~\cite{roy2017masterprint} and DeepMasterPrint ~\cite{bontrager2018deepmasterprints} sequences conduct dictionary attacks on fingerprints without knowing users' fingerprint information. Luckily, their attack success rates are very low in a high security-level setting. Specifically, the attack success rates of MasterPrint and DeepMasterPrint are only 1.88\% and 1.11\% with a 0.01\% False Acceptance Rate ($FAR$). But if the attacker has inferred some victim's fingerprint information \red1{contactlessly}, is it possible to generate stronger MasterPrint sequences for dictionary attacks?

Recently, some studies ~\cite{rathore2020sonicprint,shu2023fingersound,zhou2023FingerPattern} have shown that the uniqueness of finger-swiping friction sound is influenced by an individual's fingerprint biometric characteristics. 
Besides, such finger-swiping friction sounds can be captured by attackers online with a high possibility, which will be a catalyst to enable such attacks. 
For instance, users engage in online gaming through social applications such as Discord, where players interact and cooperate through video and in-game voice chat. In such cases, users frequently slide their fingers across the device screens. 
Similarly, during audio and video calls on mobile devices' social platforms (\eg, Skype~\cite{skype}, WeChat~\cite{wechat}, and Apple FaceTime~\cite{facetime}), users unconsciously perform swiping operations on the screen, such as searching for other apps or scrolling through information. These finger-sliding friction sounds will be transmitted to the other party by social communication software as well as malware with recording permission. 
Therefore, we will explore the possibility of an attacker remotely inferring the victim's fingerprint information based on the fingerprint friction sound.

\begin{figure}[!t]   
	\center{\includegraphics[width=8.5cm]  {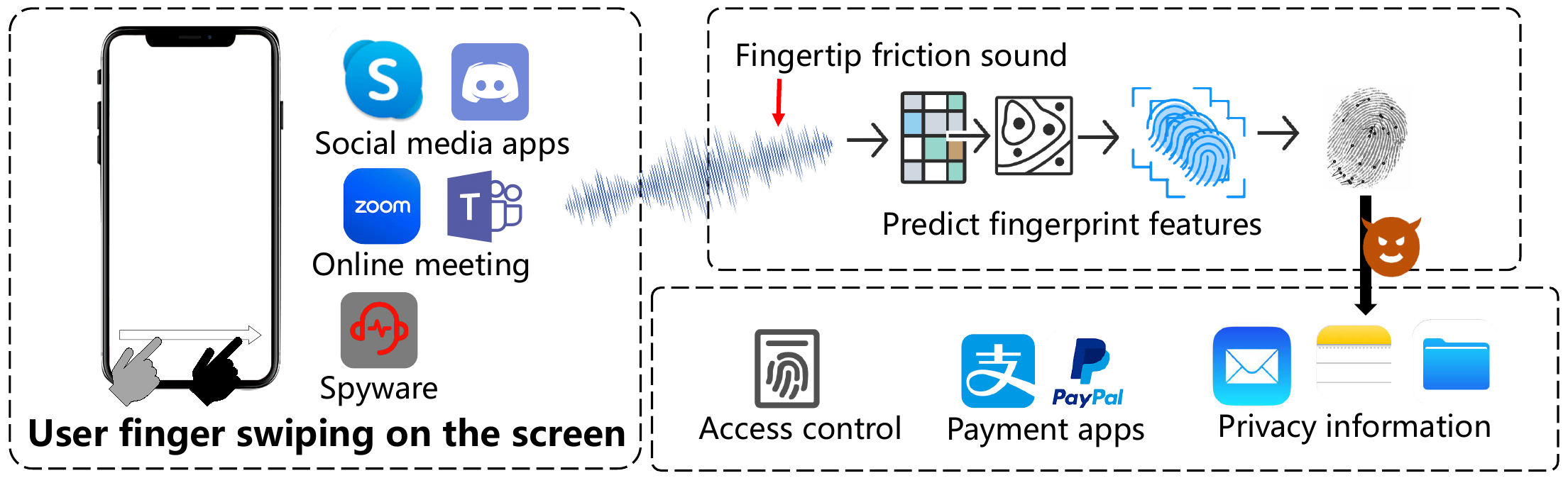}}   
	\caption{\label{Attack Scenario} Attack scenario of PrintListener.}   
\vspace{-3mm} 
\end{figure}

In this work, we propose a new side-channel attack on fingerprints, called PrintListener, which leverages users' swiping actions on the screen to extract fingerprint features and \red1{synthesize} a stronger MasterPrint sequence based on these features to conduct dictionary attacks on \red1{users'} fingerprints, as shown in Figure \ref{Attack Scenario}. To achieve this, PrintListener separates weak frictional sounds that are buried in dynamic speech and background noise interference and obtains the first-level feature (\textit{fingerprint pattern}) of fingerprints through the wide and deep combined prediction model. Further, PrintListener uses the random restart hill-climbing algorithm to synthesize the second-level feature (\textit{the position and direction of minutiae}) of fingerprints that correspond to the inferred first-level feature, namely minutiae templates, which are the basis for fingerprint authentication. In addition, the synthesized fingerprint minutiae templates can also be used to reconstruct fingerprint images~\cite{cappelli2007fingerprint}. 


By using the swiping friction sound as a natural attack entry point, our attack has two advantages:
1) Stealthiness: PrintListener can be carried out by leveraging mainstream social software with voice and video capabilities and does not necessitate any supplementary hardware. It capitalizes on the built-in microphones in electronic devices, such as smartphones, to capture the faint friction sounds generated by finger movements across electronic screens. Subsequently, the user's fingerprint patterns are inferred from these sounds.
2) Pervasiveness: PrintListener is based on the MasterPrint sequences attack, which does not require large data training on a specific person. These sequences can subsequently be employed to launch more powerful dictionary attacks on all victims' fingerprints that conform to the specific pattern.


To accomplish this goal, we need to resolve three major challenges:
1) The sound intensity of finger friction from users is extremely weak, typically ranging from 0.2 to 0.8 seconds. The original audio and video call information often contain a significant amount of redundant information. To extract the faint friction sound submerged in dynamic speech and background noise interference, we design a friction sound event localization algorithm based on spectral analysis. By moving time windows and examining the energy spectral density of audio in different frequency bands, we detect the starting and ending points of friction sound events. By analyzing the spectral peak-valley characteristics of the activity events, we can eliminate the interference from multi-band and multi-type activity noise, enabling precise localization of finger friction events. 2) Friction sound characteristics are often influenced by users' physiological and behavioral features. In addition to the primary feature patterns represented by the finger's surface morphology, there are also factors such as the pressure and velocity, as well as the swiping trajectory. To address this, we propose a joint prediction approach using a width and depth classification module to capture both interpretable audio features and deep representation features. We employ the minimum redundancy maximum relevance (mRMR) feature selection strategy and utilize conditional mutual information to reduce feature redundancy. By deploying an adaptive weighting strategy, we aim to balance the prediction results and enhance the robustness of the fingerprint feature prediction model to behavioral characteristics, thus capturing the commonalities within the fingerprint dataset with the same patterns. Furthermore, we employ pitch shift and time stretch techniques to balance the diversity of velocity and pressure in the original audio training set, including various sliding trajectories.
3) After inferring the primary pattern features of fingerprints, the potential search space for the secondary features corresponding to fingerprints of the same pattern is vast. How to effectively search and synthesize a PatternMasterPrint dictionary capable of attacking the majority of fingerprints from the latent space? To address this, we conduct a statistical analysis of the intercorrelations between the primary and secondary features and design a heuristic search algorithm specifically targeting the detailed secondary features of fingerprints. Building upon the traditional random restart hill-climbing mechanism, we first identify the crucial region (CR) within the fingerprint area, which is prone to high-frequency collisions of secondary features. Within the CR, we adaptively narrow down the potential search space based on the current search state while simultaneously increasing the likelihood of collisions among the secondary features. This approach allows us to synthesize locally optimal patterns in the master fingerprint dictionary.

Our contributions can be summarized as follows:
\begin{itemize}
	\setlength{\itemsep}{0.2pt}
	\item   
         We uncover a new side-channel attack on fingerprint and propose PrintListener, which leverages users' swiping actions on the screen to identify the fingerprint pattern and synthesizes PatternMasterPrint sequences to conduct more powerful dictionary attacks. To the best of our knowledge, this is the first work that leverages swiping friction sounds to infer fingerprint information.  
	\item 
	We design a series of algorithms for pre-processing the raw audio signals, eliminating the interference of redundant audio features, and providing a wide and deep combination prediction for fingerprint patterns. Specifically, PrintListener can automatically capture the pattern features of fingerprints from a large number of raw recordings and generate targeted synthetic PatternMasterPrints.

	
	\item 
	Extensive experimental results in real-world scenarios show that Printlistener has strong attack power on fingerprint authentication. It can attack up to 27.9\% of partial fingerprints and 9.3\% of complete fingerprints within five attempts at the highest security $FAR$ setting of 0.01\%.
	
	
\end{itemize}

\section{PRELIMINARIES}

In this section, we first analyze whether PatternMasterprint has a high probability of matching fingerprints with the same pattern, then study the feasibility of fingerprint pattern prediction via finger friction sound.

\subsection{Is PatternMasterprint's attack more powerful?}

The friction ridge details of a fingerprint are generally classified into 3 levels: Level 1 (pattern), Level 2 (minutiae points), and Level 3 (pores and ridge shape). Currently, most AFIS only use Level 1 and Level 2 features.
The Level 1 features (fingerprint pattern) can usually be divided into four distinct types: left loop, right loop, whorl, and arch~\cite{karu1996fingerprint}. Each has unique variations depending on the shape and relationship of the ridges. To determine whether PatternMasterPrint (the MasterPrint generated by fingerprints with a specific pattern) has a stronger attack power for fingerprints with the same pattern, we initially analyze the similarities between fingerprints that share the same pattern and evaluate the $FAR$ for fingerprints in datasets containing the same pattern, as well as mixed patterns. Subsequently, we investigate statistical evidence to support our intuitive hypothesis that a higher $FAR$ of fingerprints in datasets with the same pattern will increase the likelihood of detecting a PatternMasterPrint (referred to as PMP).

\textbf{Hypothesis.} The probability of finding PMP in the same pattern fingerprint dataset $ \mathcal {D}_{s}$ is $P (PMP \subset \mathcal {D}_{s})$, and the probability of finding MasterPrint (MP) in the mixed fingerprint dataset $ \mathcal {D}_{m}$ is $P (MP \subset \mathcal {D}_{m})$. Then our alternative hypothesis is
\begin{equation}
{H}_{1}: P(PMP \subset \mathcal {D}_{s}) >= P(MP \subset \mathcal {D}_{m}).
\end{equation}

\begin{figure}[!t]   
\center{\includegraphics[width=7.5cm] {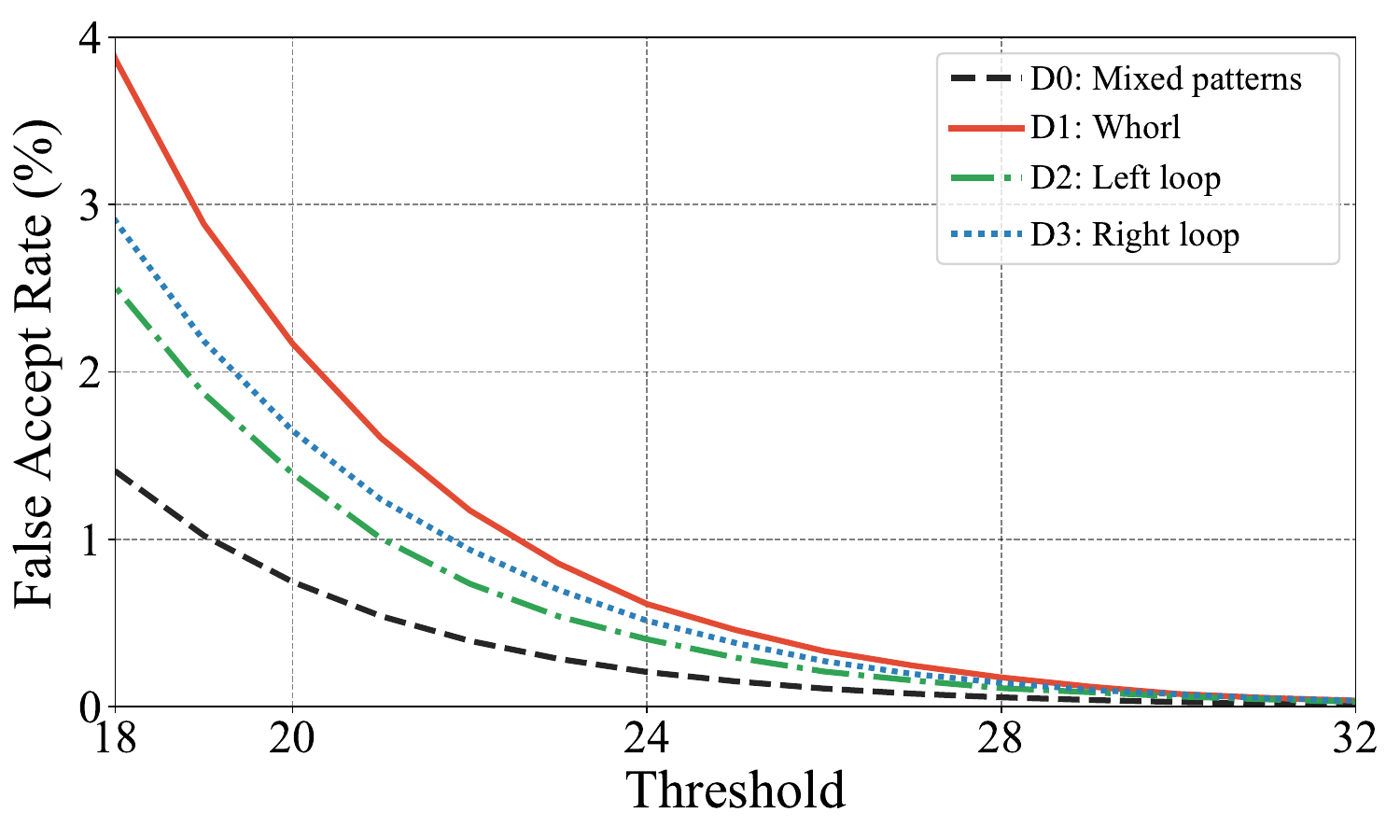}}   
\caption{\label{13-FAR-D0-D3} Variations of $FAR$ with threshold settings in D0 (mixed patterns), D1 (whorl), D2 (left loop), and D3 (right loop) Datasets.}   
\end{figure}

\textbf{Hypothesis Analysis Setup.}
We select the livedet2011\_It-aldataLive dataset~\cite{yambay2012livdet}, which comprises a total of 400 fingers, each with 5 complete fingerprint images, resulting in a total of 2000 images. It consists of 680 whorl fingerprints (dataset D1), 565 left loop fingerprints (dataset D2), 580 right loop fingerprints (dataset D3), and 175 arch fingerprints (dataset D4). Since the population of arch fingerprints is less than 5\% ~\cite{maltoni2009handbook}, we exclude dataset D4 from our statistical analysis\footnote[3]{Arch fingerprints will not be considered in this paper.}.

Thus, the mixed dataset D0 amalgamates three finger patterns ($ D0 = D1\cup D2\cup D3$). To test our hypothesis H1, the single-pattern dataset Ds can be D1, D2, or D3, and the mixed-pattern dataset Dm is D0.

\textbf{Hypothesis Statistics.}
Figure \ref{13-FAR-D0-D3} depicts the variation of $FAR$ concerning threshold values from datasets D0 to D3. With the same threshold, the $FAR$ of the single-pattern fingerprint dataset is considerably higher than that of the dataset containing a mixture of patterns. This disparity arises from the fact that fingerprints belonging to the same pattern tend to exhibit similar locations of singular points and coarse flow directions of ridges.

We define the potential MPs/PMPs as those that incorrectly match with at least 4\% of the fingerprint population. 
In the mixed fingerprint dataset D0 of 1825 fingerprints, there are 146 MPs, which accounts for an 8.00\% (146/1825) proportion of MPs. We also conducted separate statistical analyses on the D1-D3 datasets to determine the presence of potential PMPs. The proportions of PMPs in the D1-D3 datasets are as follows: 17.65\% (102/578), 25.28\% (114/451), and 13.73\% (70/510). Notably, all these proportions are higher than that of the D0 dataset, indicating a higher prevalence of PMPs in single-type fingerprint datasets. These findings provide support for the alternative hypothesis $H_1$.

\subsection{Acoustic principle of fingerprint pattern prediction}

\begin{figure}[!t]   
\center{\includegraphics[width=8.5cm]  {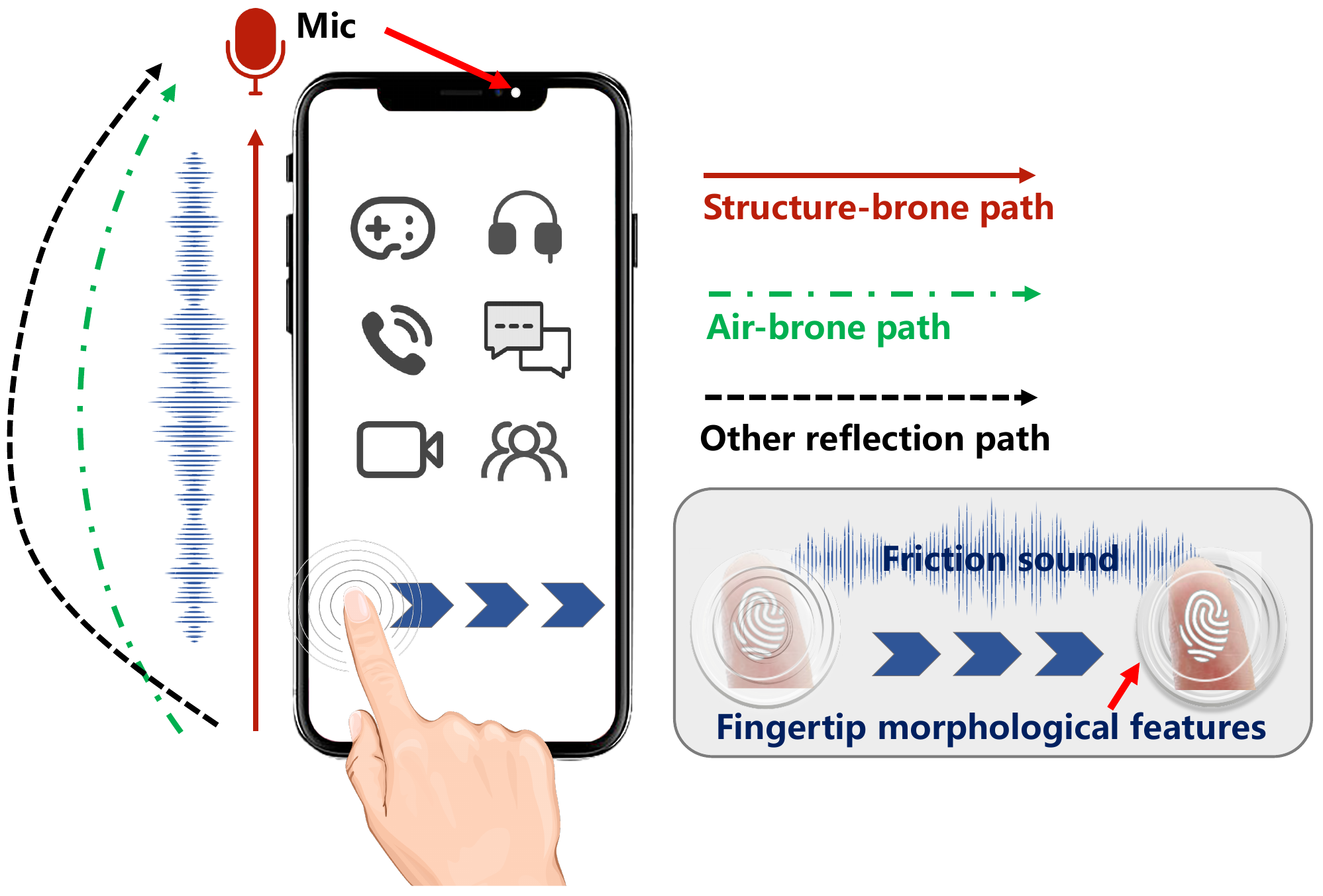}}   
\caption{\label{Sound Propagation} Propagation paths of frictional sound waves.}   
\end{figure}

\subsubsection{How are the finger frictional sounds created? }

When two solid surfaces come into contact and slide against each other, they generate vibrations and waves that result in frictional noise. Frictional noise can be divided into two categories depending on the contact pressure (high or low) between the frictional pairs~\cite{akay2002acoustics,zahouani2009friction}. Frictional noise under high contact pressure is typically transient and caused by mechanical instability between surfaces, such as door squeaks. Low-pressure frictional noise is relatively stable and is commonly referred to as roughness noise, such as the rustling sound produced by rubbing two sheets of sandpaper.

When the finger swipes on a screen, the weak coupling between the finger pad and the screen will generate a roughness noise. The production of roughness noise involves three essential factors: friction (the elastic deformation between the fingertips and the smartphone screen amplifies the vibrations), dynamics (the vibrations and waves propagate between the finger and the screen), and acoustics (audible roughness sound radiates from the finger to the surface of the phone and propagates through the air and solid medium to the phone's microphone~\cite{abdelounis2010experimental}), as shown in Figure~\ref{Sound Propagation}. 

\subsubsection{Influential factors of friction sound characteristics}

\begin{figure}[!t]   
\center{\includegraphics[width=8.5cm]  {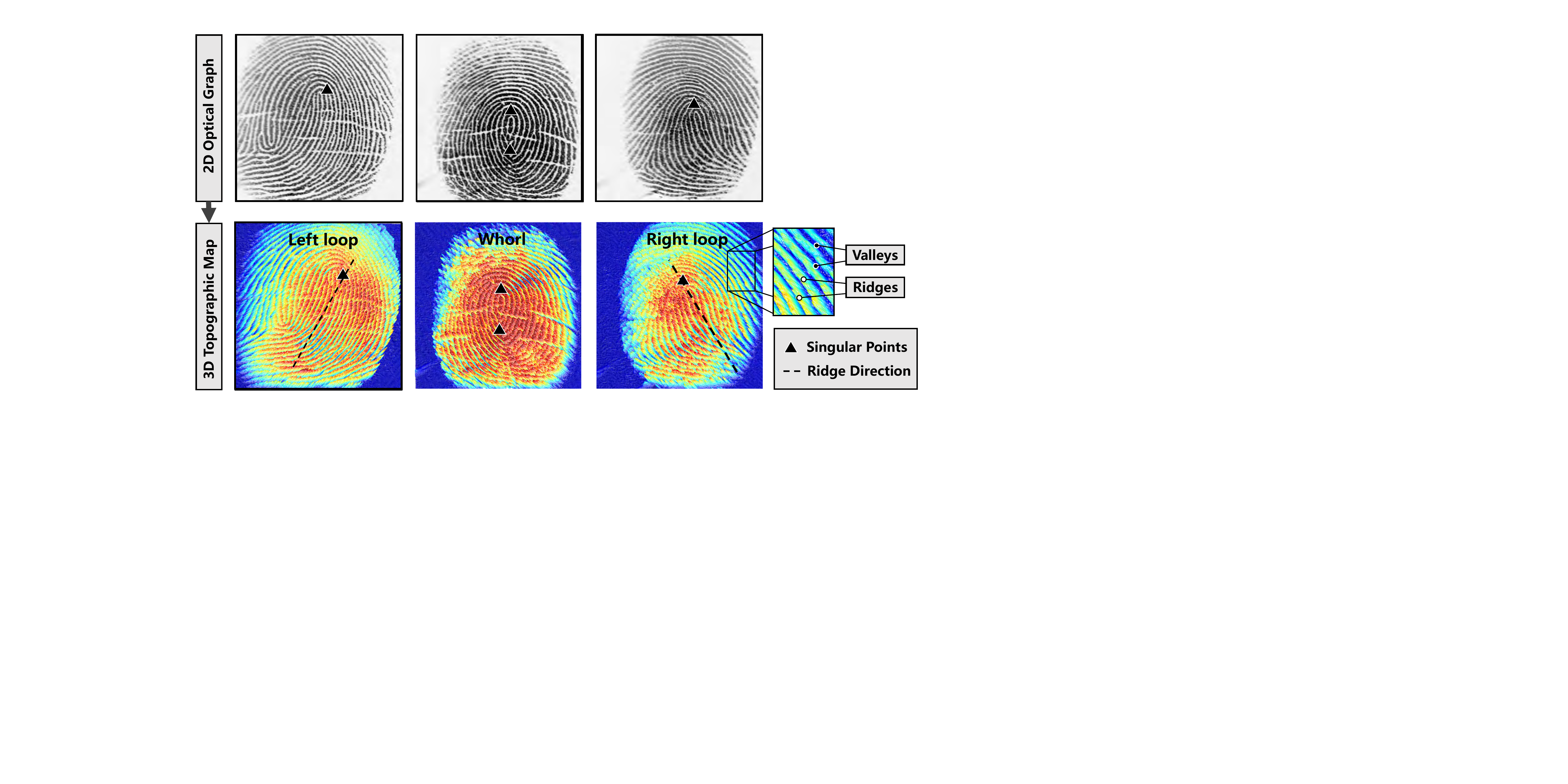}}   
\caption{\label{Figerprint Morphology} 2D optical graph and corresponding 3D topographic map (including ridge and valley lines) of the left loop, whorl, and right loop. (The 3D topographic areas of greater pressure are marked as red. The dashed lines indicate the overall direction of ridges lines, the triangles represent the singular points of different patterns).}   
\end{figure}

When the finger slides across a screen, the surface of the finger and the screen are subjected to a light load, resulting in only slight nonlinear deformation of the skin on the fingertip. The sound pressure level $L_p$ (dB) of the roughness noise can be expressed as:
\begin{equation}
L_p =20 \log _{10} {R_{a}}^{n} {V}^{m},
\end{equation}
where $R_a$ denotes the arithmetic mean value of the sliding surface roughness, $V$ represents the sliding speed, and $n$ and $m$ are two independent exponents. $R_a$ depends on the morphological features of the finger and the phone screen surface. Specifically, the surface morphology of the finger is primarily determined by the pattern of the fingerprint. Figure~\ref{Figerprint Morphology} illustrates the different fingertip morphologies, including left loop, right loop, and whorl patterns. The ridges of fingerprints reduce the contact area between the finger pad and the phone screen, resulting in variations in frictional radiation of air, solid vibration, and wave propagation modes. Intuitively, this is reflected in the characteristics of frictional sound waves~\cite{derler2015friction}. In addition to the roughness of the finger pad and the phone screen surface and the sliding speed of the finger, the characteristics of the sound waves are also influenced by the pressure applied by the finger, the humidity of the finger, and the sliding posture of users~\cite{rathore2020sonicprint}.

\subsubsection{A Feasibility Study}
Given that each user possesses unique surface morphology features on their fingerprints, we assume that the generated frictional sound waves can disclose the distinctive fingerprint pattern for certain types of mobile phone screens.

\textbf{Proof-of-concept Setup.}
We record friction sounds produced by 9 participants as they rub their fingers against the screen of the Google Pixel 4, which is covered with a matte screen protector. In the first experiment, participants gradually increase the pressure and speed of their finger rubbing while swiping their fingers across the smartphone screen 15 times. In the second experiment, participants wrap their fingers with transparent tape and repeat the same rubbing motion as in the first experiment. Our objective is to identify specific acoustic features in the friction sounds that correspond to different fingerprint ridge patterns.

\begin{figure}[!t]   
    \begin{minipage}{.25\textwidth}
    \centering
    \subfloat[]{\label{Proof-of-concept a}\includegraphics[width=1\textwidth]{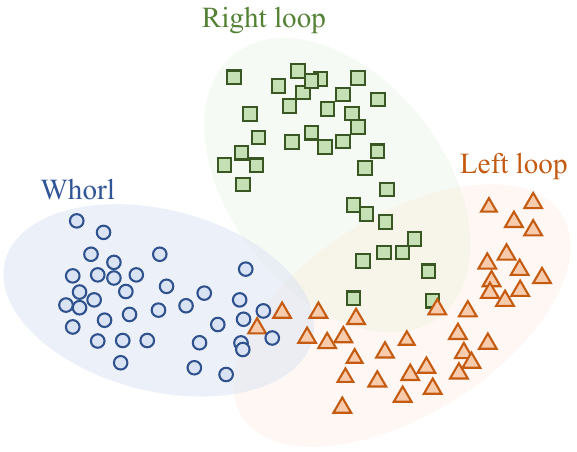}}
    \end{minipage}
    \hfill
    \begin{minipage}{.19\textwidth}
    \centering
    \subfloat[]{\label{Proof-of-concept b}\includegraphics[width=1\textwidth]{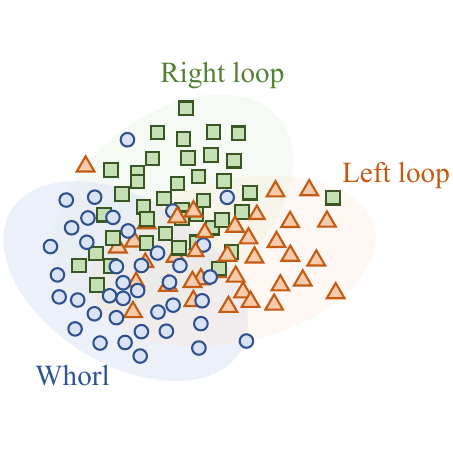}}
    \end{minipage}
\caption{A proof of concept for two-dimensional mapping of frictional sound characteristics of (a) different fingerprint patterns; (b) blurred fingers.}
\label{Proof-of-concept}
\end{figure}

\textbf{Result.}
To compare the characteristics of friction sounds, Figure \ref{Proof-of-concept} presents the two-dimensional mapping of frictional sound characteristics of Mel frequency cepstral coefficients (MFCC), Mel spectrograms, and other frequency-domain and time-domain features. Notably, each data point of the friction sound formed a cluster corresponding to a unique fingerprint pattern. As shown in Figure \ref{Proof-of-concept a}, under the influence of pressure and velocity, different fingerprint patterns can still be roughly distinguished with overlap. However, the corresponding features cannot be distinguished in Figure \ref{Proof-of-concept b} when the fingers are blurred. Thus, the fingerprint pattern depends on the fingerprint itself rather than the shape of the fingertip set or the contact area.

Based on the friction sounds, features such as MFCC can be utilized to distinguish corresponding fingerprint patterns within a small range. Hereafter, we aim to enhance the accuracy of fingerprint feature prediction.


\section{Threat Model}


\textbf{Goal.} 
The attacker aims to leverage PrintListener to deduce extensive fingerprint information of users by analyzing the friction sound produced when users swipe on their phone screens while engaged in audio and video calls on social media platforms. For instance, users commonly swipe their fingers to scroll through the browse messages or news during audio and video calls, or frequently perform swipe operations when engaging in online gaming through social applications.
In this paper, we focus on generating targeted synthetic PatternMasterPrints with the following assumptions.

\begin{table} [!t]
\renewcommand{\arraystretch}{1.1}
  \caption{\red1{Guessing framework of PrintListener.}}
  \centering
  \label{tab: guessing_framework}
  \normalsize
  \renewcommand\arraystretch{1.1}
  \begin{tabular}{c|c}
    \hline
      Guessing path & Example \\    
    \hline
      Instant audio/video & Skype / Wecom / Facetime\\
     \hline 
      Online meetings & Zoom / Google Meet\\
     \hline 
      Live streaming & Twitch / Youtube Live / Discord\\
    \hline 
      Others & Malware / Swipe keyboard\\     
    \hline 
    
  \end{tabular}
\end{table}

\begin{figure*}[!t]   
\center{\includegraphics[width=18cm]  {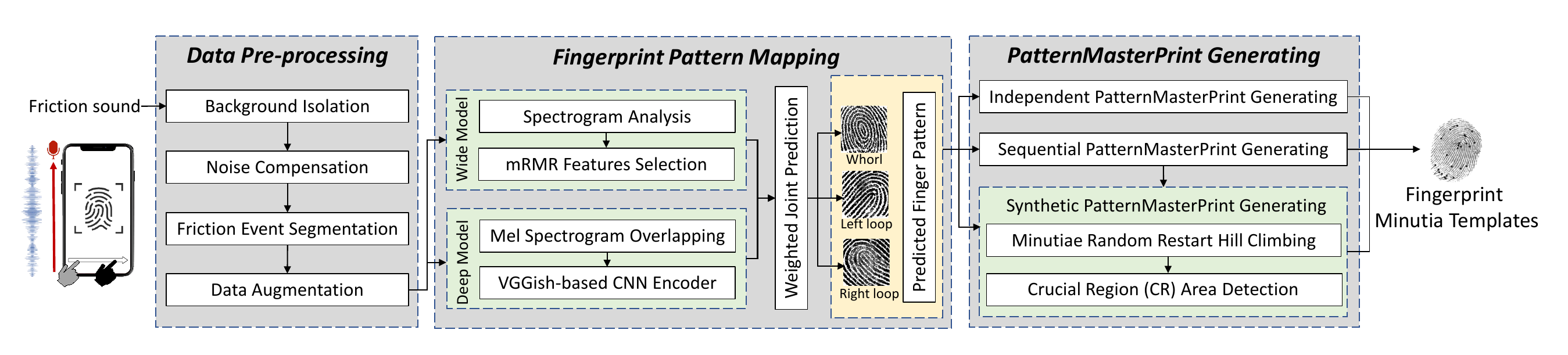}}   
\caption{\label{System Overview} The workflow of PrintListener.}   
\end{figure*}

\textbf{Capabilities and Knowledge.}
We assume the attacker can access the victim's finger friction sound. For example, the attacker has the victim's contact information, which could include the victim's phone number or account details for popular audio and video social software platforms (\eg, Google Meet, Skype, Discord, \etc). Then, the attacker may proceed to initiate a voice or video call with the victim, connect to voice collaboration software for playing games, or infiltrate the same online voice or video conference as the victim. Malware with recording permissions can also silently record the swiping friction sound. This assumption is close to the practical scenario.
\red1{The attacker's guessing framework is illustrated in Table~\ref{tab: guessing_framework}.
}



\textbf{No adversary proximity/accessibility.} We assume the user is cautious about the surrounding environment and protects the fingerprint carefully. The user will not actively disclose the fingerprint to others nearby \red1{or casually leave fingerprint impressions on personal or public devices. Thus, attackers are unable to obtain the fingerprint pattern information through observation or proximity to the devices used by the user}.

\textbf{No access to personal devices.} The attacker cannot obtain any private devices (such as mobile phones, tablets, and other electronic devices with interactive screens) touched by the user. \red1{Once the PatternMasterPrints for the victim are created, their efficacy can be tested on public devices that use the user's fingerprint, such as fingerprint-operated public devices used for time attendance and access control. Besides, the attackers can also directly sell generated candidate fingerprints to other malicious parties, which will also cause a significant loss to users.}


\textbf{No additional user action.} Attackers cannot require any additional actions from the user, such as asking the user to rub their finger pads on the screen repeatedly. Throughout the entire attack process, the attacker should ensure that the user is unaware and only collects the sounds stealthily during regular audio and video calls. 


\section{Design of PrintListener}

\subsection{Overview}

After eavesdropping on the user's finger friction sound through a social network, PrintListener generates a specialized PatternMasterprint sequence specifically designed for the user's fingerprint. As shown in Figure~\ref{System Overview}, the workflow consists of three modules: \textit{Data Pre-processing}, \textit{Fingerprint Pattern Mapping}, and \textit{PatternMasterPrint Synthesizing}. 

1) Data Pre-processing: This module separates weak friction sounds that are buried in dynamic speech and background noise interference. High-pass filters are utilized to eliminate low-frequency noise, and a spectrum density analysis method is employed to obtain precise segments of the friction sound. Finally, waveform resampling is applied to extend the frictional segments and optimize the original dataset.

2) Fingerprint Pattern Mapping: This module employs width and depth combination models to extract interpretable audio and deep representation features from the augmented friction sound segments. An adaptive weighting strategy is employed to balance the predicted results of the primary feature pattern. 

3) PatternMasterPrint Generating: This module first samples independent and sequential PatternMasterPrints (PMPs). Based on the sequential PMPs, a random restart hill-climbing algorithm based on a crucial region synthesizes a PMP dictionary in the secondary feature latent space. The synthetic PMP dictionary is capable of high-frequency collision with fingerprints of the same pattern. 

\begin{figure*}[!t]   
\center{\includegraphics[width=18cm]  {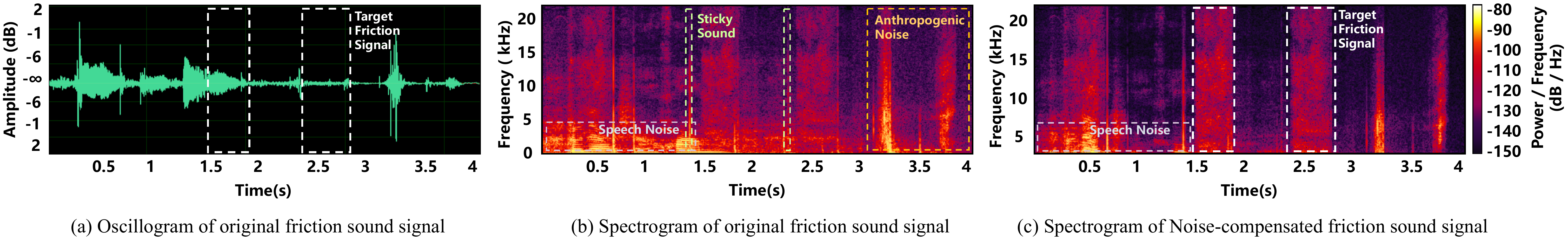}}   
\caption{\label{Sound_Wave_Diagram} \red1{Data pre-processing before friction event segmentation.}}   
\end{figure*}



\subsection{Data Pre-processing}

Once the raw audio and video call voice are obtained, PrintListener removes noise and compensates for the target frictional sound signal. Subsequently, it obtains precise segments of the frictional sound based on the frequency spectrum and finally performs data augmentation on the acquired frictional sound segments.

\subsubsection{Background Noise Isolation}
The intensity of finger friction sound is generally low, and it is frequently subject to interference from either steady background white noise or non-stationary additive noise (\eg, dynamic speech or device electronic noise). When we use mobile phones for audio and video calls in indoor environments, such as family rooms and offices, the noise energy is typically concentrated below 4 kHz. Environmental noise can be disregarded when the frequency range exceeds 8 kHz~\cite{zhou2018dolphin}. To eliminate low-frequency noise while preserving the fingerprint information, we select a finite impulse response (FIR) high-pass filter~\cite{neuvo1984interpolated} with a 4 kHz passband to erase the background noise.

\subsubsection{Noise Compensation}
The method based on short-time amplitude and phase spectrum noise compensation~\cite{islam2018speech} is widely used to enhance the target signal degraded by additive noise without introducing any distortion.
The speech signal is analyzed frame by frame, and the noisy speech is segmented into overlapping frames by sliding windows. Under the condition that the noise speech phase remains unchanged, the corrected complex spectrum is obtained by aggregating the corrected amplitude spectrum.

\begin{equation}
\left|Z^{t}\left[\omega_{k}\right]\right|=\left|G_{G A}\left[\omega_{k}\right] \cdot Y^{t}\left[\omega_{k}\right]\right|,
\end{equation}
where $G_{G A}\left[\omega_{k}\right]$ represents the gain function of the conventional geometric approach. $Y^{t}[\omega_{k}]$ is the Fast Fourier Transform (FFT) representation of it. Enhanced speech frames $\hat{x}^{\tau}[n]$ are synthesized by performing an Inverse Fast Fourier Transform (IFFT) on the corrected cepstral coefficients $Z^{\tau}[\omega_{k}]$.

\begin{equation}
    \hat{x}^{\tau}[n]=\operatorname{Re}\left(\operatorname{IFFT}\left\{\left|Z^{\tau}\left[\omega_{k}\right]\right| f\left(Z^{\tau}\left[\omega_{k}\right]+\varphi^{\tau}\left[\omega_{k}\right]\right)\right\}\right),
\end{equation}

where Re(·) denotes the real part of a complex number,  $\tau$ represents the frame number,  $Z^{\tau}[\omega_{k}]$ is the modified phase of $Z^{t}[\omega_{k}]$, and $f(Z^{\tau}[\omega_{k}]+\varphi^{\tau}[\omega_{k}])$ represents the corrected complex spectrum obtained by adding the corrected phase spectrum $Z^{\tau}[\omega_{k}]$ and the corrected amplitude spectrum $\varphi^{\tau}[\omega_{k}]$. The final enhanced speech signal is synthesized using the standard overlap and add method.

\subsubsection{Friction Event Segmentation}

To accurately extract specific friction sound segments from recordings and exclude redundant information about friction segments, we design a friction sound event localization algorithm based on the frequency spectrogram of audio signals. By analyzing the energy spectrum density changes across the entire frequency range, we perform friction sound event localization. Firstly, we obtain the spectrogram of the original audio signal by applying the Fourier transform. Then, for each time window, we calculate the squared amplitude of the frequency-domain signal of the audio frame, to obtain the power spectral density of the original audio at different frequencies. Next, we conduct a three-step detection on the entire activity signal:

\textbf{Step 1: Silent regions exclusion.}
When friction sound or activity noise (such as continuous human speech) occurs or ends, the difference in energy density between adjacent time windows increases, which is reflected in the spectrogram as a sudden brightening or darkening of color. 
During silent periods, the spectral energy is weaker, and the difference in spectral energy between adjacent time windows is lower.

The audio energy spectral density $P(n, k)$ of the $k$-th frequency component at the $n$-th sample point in the time window T is calculated as:

\begin{equation}
P(n, k) = \left| \sum_{m=0}^{N-1} w(m) \cdot x(n+m) \cdot e^{-j2\pi km/N} \right|^2{,}
\label{con:stft}
\end{equation}

where $N$ denotes the number of samples in each time window, $w(m)$ is the window function, and $x(n+m)$ denotes the $(n+m)$-th sample point in the time window. This equation uses the window function to smooth the data within the time window and then performs a short-time Fourier transform to obtain the energy spectral density of the different frequency components. By moving the time window and comparing the energy differences between adjacent time windows, we first exclude the silent regions.

\textbf{Step 2: Full-frequency energy verification.} Through extensive experimentation, we have observed that the spectral energy of friction sounds uniformly increases across the entire frequency range of $0\sim22$ kHz. However, common interfering sounds, such as human speech are usually distributed in low-frequency bands. Therefore, we set a threshold $T_{Dif}$ for the difference in spectral energy density between adjacent time windows and a threshold $T_{Var}$ for the variance of spectral energy density across $0\sim22$ kHz of the same time window, to analyze the potential friction events.
When the difference in spectral energy density between adjacent time windows is greater than $T_{Dif}$, and the variance is smaller than $T_{Var}$, it is considered a potential starting or ending point of friction sound segments.

\textbf{Step 3: Duration verification.}
After obtaining the potential friction events, we verify the duration of these activity segments. Only when the duration of the activity noise exceeds $T_{Dur}$, it is considered as the friction sound. Through extensive testing, we adjust the values of $T_{Dif}$, $T_{Var}$, and $T_{Dur}$ to accurately locate friction sound events. As shown in Figure~\ref{Sound_Wave_Diagram}, most friction sound fragments can be automatically segmented out.

\begin{figure}[!t]   
    \center{\includegraphics[width=0.44\textwidth]{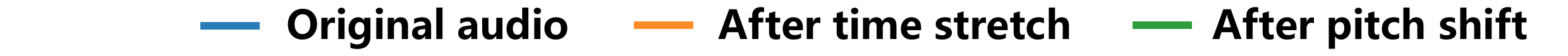}}
    
    \begin{minipage}{.052\textwidth}
        \centering
        \includegraphics[width=1\textwidth]{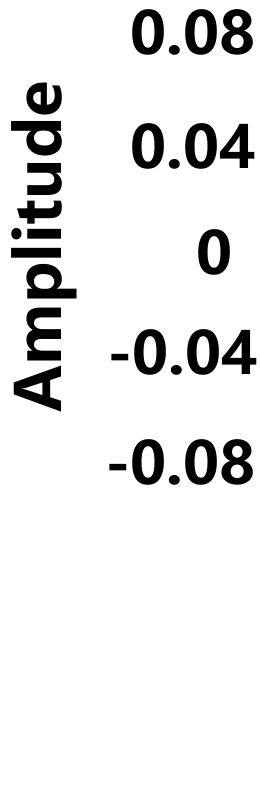}
    \end{minipage}
    \begin{minipage}{.2\textwidth}
        \centering
        \subfloat[rate = 0.8]{\label{da_a}\includegraphics[width=1\textwidth]{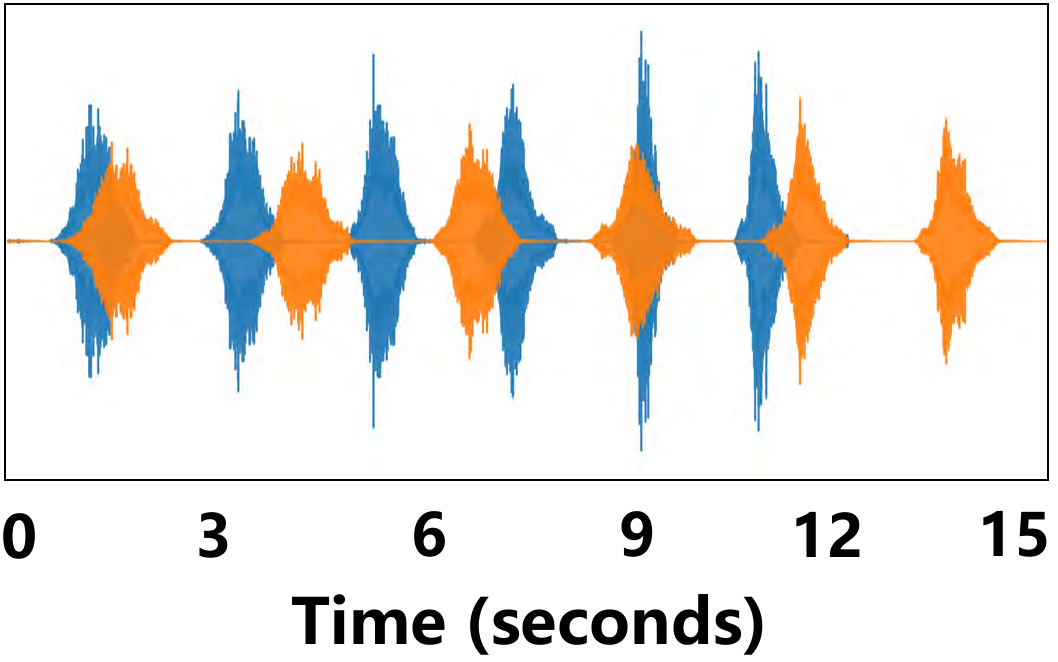}}
    \end{minipage}
    \begin{minipage}{.203\textwidth}
        \centering
        \subfloat[rate = 1.2]{\label{da_b}\includegraphics[width=1\textwidth]{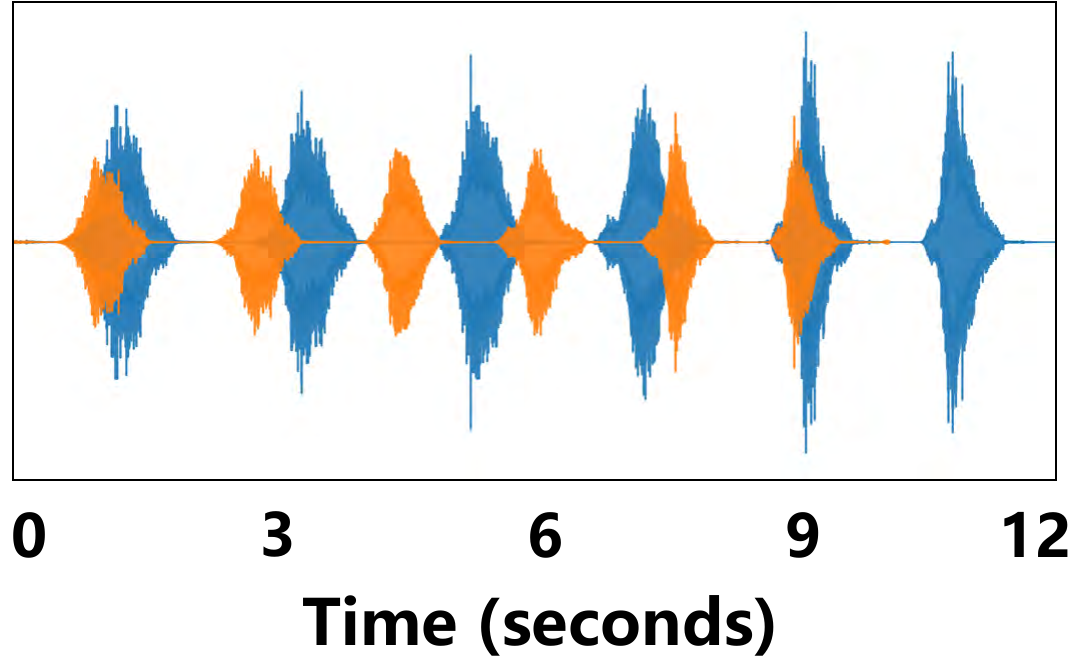}}
    \end{minipage}

    \begin{minipage}{.052\textwidth}
        \centering
        \includegraphics[width=1\textwidth]{data_augmentation_label.pdf}
    \end{minipage}
    \begin{minipage}{.2\textwidth}
        \centering
        \subfloat[n\_steps = 2]{\label{da_c}\includegraphics[width=1\textwidth]{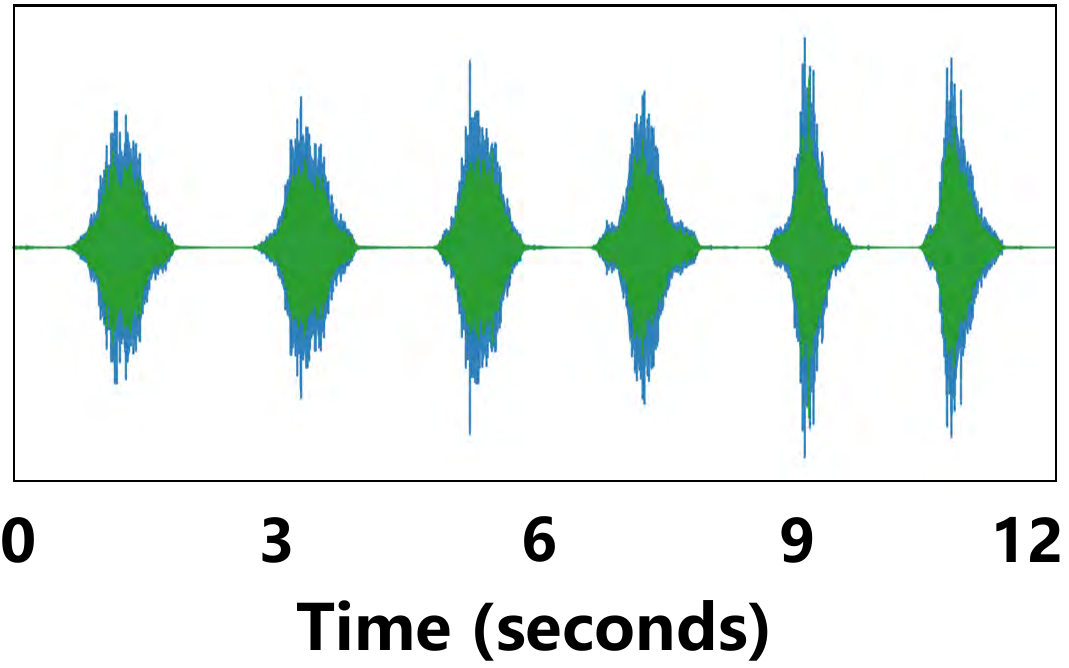}}
    \end{minipage}
    \begin{minipage}{.2\textwidth}
        \centering
        \subfloat[n\_steps = -2]{\label{da_d}\includegraphics[width=1\textwidth]{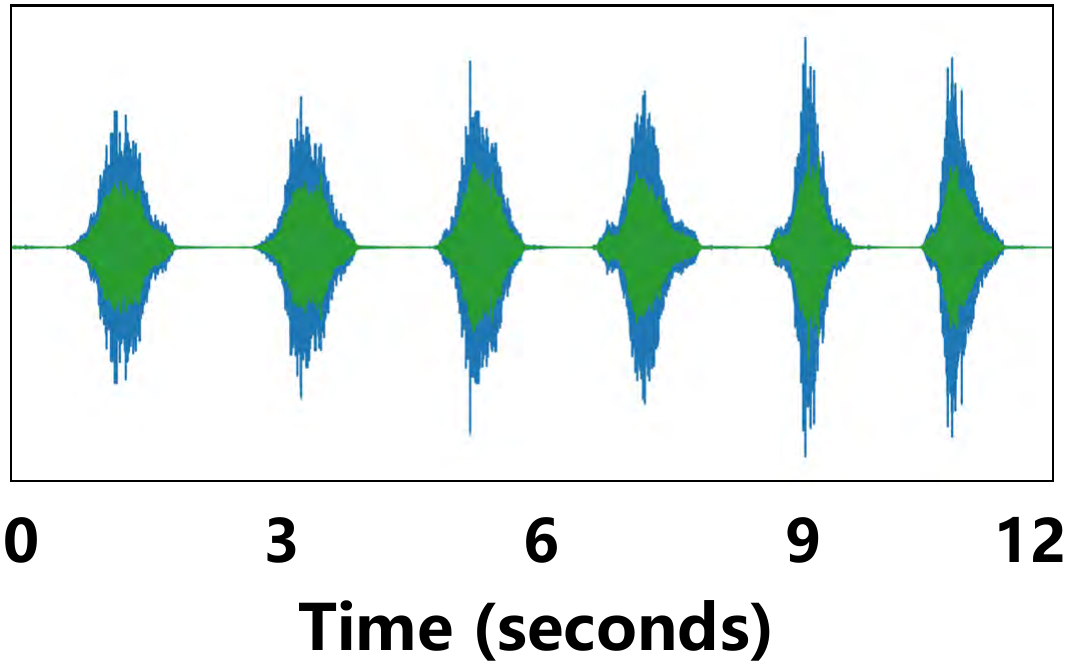}}
    \end{minipage}

\caption{Data augmentation}
\end{figure}

\subsubsection{Data Augmentation}
Waveform resampling is a common data augmentation technique. In our study, we use time stretching and pitch shifting~\cite{nanni2020data} to generate a range of similar yet distinct waveforms. This expands our datasets and improves the physical reality of samples (\eg different swiping pressures and speeds). 

\red1{
\textbf{Time Stretch.}
The user's swiping speed can interfere with the features of acoustic signals. PrintListener employs the waveform similarity overlap-and-add (WSOLA) algorithm \cite{verhelst1993overlap} to modify the temporal scale of the signals. The original waveforms are accelerated by a factor of 0.8 and 1.2 respectively (As shown in Figure \ref{da_a} and \ref{da_b}). }

\red1{
\textbf{Pitch Shift.}
In the actual attack scenario, the configuration and orientation of the microphone, the touchscreen material, and the state of the user's finger (dry or wet) can vary. Therefore, even if the finger slides along the phone screen in the same direction, the collected friction sound will have a different pitch trajectory.
To enhance the training data, PrintListener utilizes pitch shifting to slightly alter the pitch of the frictional sound waveforms, increasing or decreasing it by 2 semitones (As shown in Figure \ref{da_c} and \ref{da_d}).
}

\subsection{Fingerprint Pattern Mapping}

After pre-processing the raw friction sound, PrintListener utilizes wide models to extract interpretable audio features and VGG-based deep models to extract deep representation features, then employs an adaptive weighting strategy to balance the predicted results.

\subsubsection{Interpretable Audio Features Extraction}

We use a two-stage features selection strategy and analyze the interpretable audio features comprising six frequency domain features and three cepstral domain features. 

\begin{table} [!t]
  \renewcommand{\arraystretch}{1.1}
  \caption{Selected interpretable audio features.}
  \centering
  \label{tab: features}
  \normalsize
  \begin{tabular}{cllc}
    \hline
    Domain & Feature & Feature vector \\
    \hline
    \multirow{6}{*}{\rotatebox{90}{Frequency}} &  LSF & $f1-f12$ \\
       & Chroma  & $f13-f24$ \\
       & Spectral Kurtosis & $f25$\\ 
       & Spectral Skewness & $f26$\\
       & Spectral Contrast & $f27-f33$ \\
       & Spectral Centroid & $ f34 $ \\
    \hdashline
    \rule{0pt}{10pt}
    \multirow{3}{*}{\rotatebox{90}{Cepstral}} & MFCC &  $f35-f73$ \\
      & LPCC & $f74-f86$ \\
      & RASTA-PLP & $f87-f99$ \\

    \hline
\end{tabular}
\end{table}

\textbf{Spectrogram Analysis.} Since the important perceptual characteristics of the audio signal are in the power spectrum, we need to perform a frequency domain analysis. Firstly, the friction sound is processed with framing and windowing to prevent the spectral leakage of the signal.
Next, the discrete Fourier transform (DFT) is employed to convert the time-domain signal to the frequency-domain signal. 
We obtain the cepstrum coefficients for each frame of the frequency-domain signal by performing a DCT transform on the logarithmic amplitude of the DFT.

For each frame of the frequency-domain signal, we obtain the cepstrum coefficients by performing a DCT transform on the logarithmic amplitude of the DFT.

\textbf{mRMR Feature Selection.} After extracting common frequency-domain and cepstral-domain features, we adopt the minimum redundancy maximum relevance (mRMR)~\cite{peng2005feature} criterion and use mutual information to select features. 
Firstly, we perform a first-order incremental search to maximize the mutual information between the selected features and the class variable, thereby identifying a candidate feature set in the feature space. Since mRMR only considers local optima further to weigh the correlation and redundancy between different features, we perform a second feature selection in the way of Wrapper until the classification accuracy and feature dimensionality reduction are not improved.
This process allows for the extraction of six frequency-domain features and three cepstral-domain features: linear spectral frequency (LSF), chroma, spectral kurtosis, spectral skewness, spectral contrast, spectral centroid,  Mel-frequency cepstral coefficients (MFCC), linear prediction cepstral coefficients (LPCC), and RASTA-PLP. The dimensionalities of feature vectors are represented in Table~\ref{tab: features}.

\subsubsection{Deep Representation Features Extraction}
In this paper, we leverage a VGGish-based model for audio analysis, which has a powerful audio feature learning capability.
We employ the feature-based transfer learning method to fine-tune the pre-trained VGGish-base model. Our VGG-like network is trained to learn feature mappings from the high-dimensional space of raw friction sound to the corresponding fingerprint patterns. The implementation details of the VGGish-based CNN Encoder are explained in APPENDIX A.

\textbf{Mel Spectrogram Overlapping.} We process the friction segment in accordance with the input requirements of the model network. Firstly, we calculate the one-sided short-time Fourier transform (STFT) of the friction sound segment with a 20 ms periodic Hann window and a 10 ms non-overlapping shift. The spectral density is computed by Formula~\ref{con:stft}. Secondly, we map the resulting spectrogram onto a 64-band Mel filter bank to calculate the Mel spectrogram, which is buffered into 96 overlapping spectrograms. 
Thus, a $96 \times 64\times1$ matrix represents each friction sound segment.

\textbf{VGGish-based CNN Encoder.} We use a VGG-like model trained on AudioSet, with $96 \times 64 \times 1$ of friction sound segment as input and a semantically meaningful $1 \times 128$ dimensional embedding as output, which is sent to the downstream classification model. The deep network consists of multiple layers of filters that capture different levels of features in the sound, including 8 convolutional layers, 5 pooling layers, and 2 fully connected layers. 

\subsubsection{Weighted Joint Prediction}

To minimize overfitting and accurately assess classification performance, we divide the dataset into training, validation, and testing sets in a 6:2:2 ratio and employ 10-fold cross-validation. We assessed the classification performance of four widely-used multi-classifiers: K-Nearest Neighbors algorithm (KNN), Decision Tree, Random Forest, and Adaboost. Meanwhile, due to traditional machine learning models' high interpretability but relatively low classification accuracy, PrintListener combines a CNN network to calculate the joint distribution of features.

We employ three segments of friction sound to predict the fingerprint pattern for each finger. The labels $i$ can be 1, 2, or 3 and correspond to the left loop, right loop, or whorl pattern, respectively. The predicted $score_{i}$ for each pattern is calculated as follows:

\begin{equation}
score_i = \sum_{j=1}^{3}  {w}_{wide} \cdot prew_{ji} + {w}_{deep} \cdot pred_{ji},
\end{equation}

where $j$ represents the jth segment of friction sound used for prediction, ${w}_{wide}$ / ${w}_{deep}$ denote the weights assigned to two modules, and $prew_{ji}$ / $pred_{ji}$ indicates whether the jth segment of friction sound is predicted as the ith fingerprint pattern. If the jth segment is predicted as the ith pattern, $prew_{ji} / pred_{ji} = 1$, and otherwise, $prew_{ji} / pred_{ji} = 0$. Corresponding weights are set for the two classification modules through voting. From empirical analysis, the combination of the KNN classifier and VGGish-like network is found to be optimal. The final predicted fingerprint pattern corresponds to the label with the highest value among $score_{1}$, $score_{2}$, and $score_{3}$.

\subsection{PatternMasterPrint Synthesizing}
This section introduces three methods for generating PatternMasterPrint: independent PatternMasterPrints generating, sequential PatternMasterPrints generating, and synthesized PatternMasterPrints generating. The first two methods involve direct sampling from existing fingerprint datasets, while the synthesized PatternMasterprints are generated by using a random restart hill-climbing algorithm applied to a large dataset of fingerprints with diverse patterns. This approach enables them to adapt to various image qualities and features. These three types of PatternMasterprints are designed explicitly for fingerprint authentication systems that rely on minutiae templates.
Given that conventional smartphone fingerprint authentication systems often permit up to 5 attempts~\cite{touchid}, PatternMasterPrints are composed of five fingerprints intended to be used consecutively to compromise the authentication system.

\subsubsection{Independent PatternMasterprint Generating}
For a dataset containing n left-loop fingerprint images, we set it as dataset A and perform exhaustive matching on all fingerprints in the dataset to identify the corresponding independent PatternMasterPrint sequence.
For each fingerprint image ${x}_{i}$ in dataset A, we perform pairwise matching between ${x}_{i}$ and all fingerprint images ${x}_{j}$ in dataset A to obtain the matching score ${M}_{i}$, which is represented as follows:

\begin{equation}
{M}_{i}=\sum_{j=1}^{n} p({x}_{i},{x}_{j}) / n{,}
\label{con:cal_Mi}
\end{equation}

\begin{equation}
	p({x}_{i},{x}_{j}) = 
    \begin{cases}
    	1, & score({x}_{i},{x}_{j})> \Lambda, \\
    	0, & otherwise{,}
	\end{cases}
\label{con:pxij}
\end{equation}

where $\Lambda$ represents the matching threshold, $n$ is the number of fingerprint images in dataset A.

We sort ${M}_{i}$ in descending order and select the fingerprint images corresponding to the top-five ${M}_{i}$ as independent left-loop PatternMasterPrints.
Sampling independent PatternMasterPrints from a fingerprint dataset offers the advantage of computational simplicity and speed. However, there can be some overlap in the set of victim fingerprints targeted by different independent PatternMasterPrints. To address this issue, we introduce the second method to obtain sequential PatternMasterPrints. The process is the same for the right loop and whorl. In the following, we also take left-loop PatternMasterPrints generating as an example.

\subsubsection{Sequential PatternMasterPrint Generating}\label{sec:sec5.2}
The sequential PatternMasterPrints are generated similarly to the independent PatternMasterPrints but require 5 rounds of computation. First, for each fingerprint image ${x}_{i}$ in dataset A, the matching score ${M}_{i}$ is calculated by Equations \ref{con:cal_Mi} and \ref{con:pxij}. The fingerprint image corresponding to the highest ${M}_{i}$ value is selected as the first sequential left-loop PatternMasterPrint. This fingerprint image is then removed from dataset A, and the previous steps are repeated for each remaining fingerprint image to obtain the second sequential PatternMasterPrint. This process is repeated five times to yield five sequential PatternMasterPrints.

As both independent PatternMasterPrints and sequential PatternMasterPrints are derived from real user fingerprint images, the randomness of features (\eg the size of the fingerprint area, the position or direction of minutiae) is low. This will reduce their matching ability when attacking other fingerprint images. 
Moreover, finding PatternMasterPrints with a high impersonation rate in selected fingerprint image datasets may not always be possible. To address these issues, we propose a random restart Hill-climbing algorithm for synthesizing PatternMasterPrints artificially and improving their generalization ability.



\begin{figure}[!t]   
\center{\includegraphics[width=7.5cm] {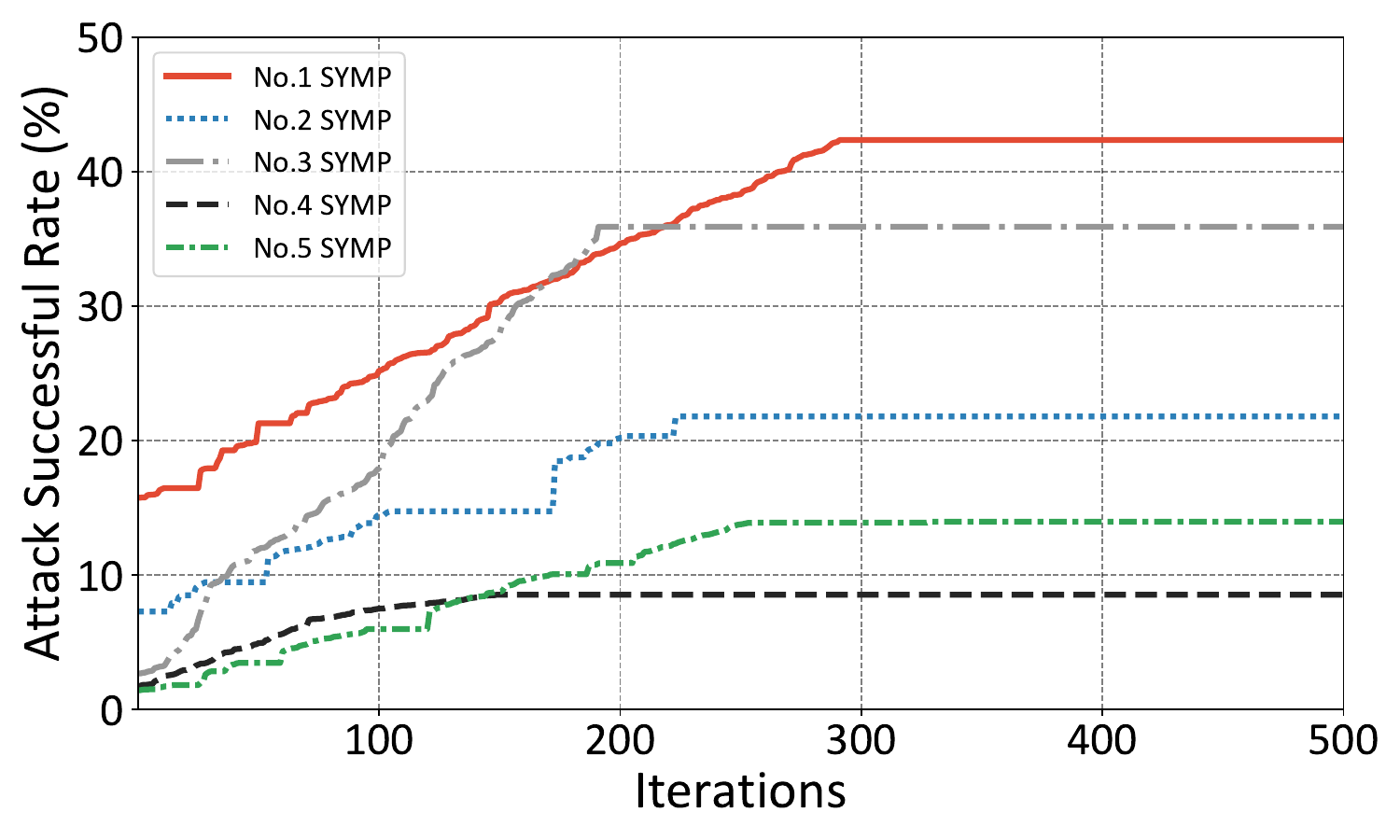}}   
\caption{\label{12-Hill-Climbing} \red1{Variation of attack successful rate with iterations of hill-climbing algorithm.}}   
\end{figure}

\begin{table*} [!t]
\renewcommand{\arraystretch}{1.1}
  \caption{\red1{Thresholds of three fingerprint datasets at different false acceptance rates ($FAR$).}}
  \centering
  \label{tab: Threshold and TAR}
  \footnotesize
  \normalsize
  \begin{tabular}{| p{2.2cm}<{\centering} | p{0.9cm}<{\centering} p{0.9cm}<{\centering} p{0.9cm}<{\centering} | p{0.9cm}<{\centering} p{0.9cm}<{\centering} p{0.9cm}<{\centering} | p{0.9cm}<{\centering} p{0.9cm}<{\centering} p{0.9cm}<{\centering} |}
    \hline
      Datasets & \multicolumn{3}{c|}{FingerPassDB7} & \multicolumn{3}{c|}{Livedet2011 ItalData} & \multicolumn{3}{c|}{PatternFinger} \\
    \hline   
      $FAR$ (\%) & 1 & 0.1 & 0.01 & 1 & 0.1 & 0.01 & 1 & 0.1 & 0.01 \\
    \hline
      Threshold & 12 & 19 & 28 & 18 & 25 & 31 & 23 & 31 & 41 \\
    \hline
     
\end{tabular}
\end{table*}

\subsubsection{Synthetic PatternMasterPrint Generating}
In this section, we iterate on the previously generated sequential PatternMasterPrints using a random-restart hill-climbing algorithm to obtain synthetic PatternMasterPrints. To improve the efficiency of the iteration process, we detect the central region (CR) area in the fingerprints.

\textbf{Crucial Region Area Detection.}
The detailed template of a fingerprint is determined by the details and orientation of each secondary feature of the fingerprint, such as ridge endings and bifurcation points. The seed fingerprint is divided into multiple units. To improve the efficiency of the iteration, we analyze the central region (CR) of fingerprints, which is the area with a high probability of fingerprint minutiae collision in the fingerprint image and is usually located in the middle region. We first define a middle area of 200 x 250 pixels as a crucial region and then conduct the random restart hill-climbing algorithm iteration within the CR area. To prevent unnecessary iterations, each unit covers a ridge spacing on the fingerprint (\eg, 9 pixels on a 500 dpi fingerprint image). Meanwhile, PrintListener quantizes the range of the fine direction into 16 equidistant intervals, each representing a change of 22.5 degrees.

\textbf{Minutiae Random Restart Hill Climbing.}
\red1{The specific iterative algorithm aims to obtain the synthetic PatternMasterPrint template by searching within the training set and maximizing the Attack Success Rate (ASR).} Due to the difficulty of searching the entire space, this paper adopts a local search approach to synthesize the PatternMasterPrint. Specifically, a random-restart hill-climbing algorithm is employed to synthesize the PatternMasterPrint. The initial seed for each round is the sequential PatternMasterPrint obtained in Section~\ref{sec:sec5.2}. In each round, the details of the template are incrementally or degressively modified to create a new PatternMasterPrint and iterative modifications are made based on the output scores of the matcher. Throughout the iterative process, the best-performing detail template serves as the stored state. If the new iteration of hill-climbing produces a template  $fp_x$ that is better than any previous generation, it replaces the existing stored state. The iteration stops when  $fp_x$ meets the given Attack Success Rate or the hill-climbing iterations reach the predefined maximum value. 
 \red1{The details of the random restart hill-climbing algorithm~\ref{alg:Hill-Climbing} are described in APPENDIX.}

Figure~\ref{12-Hill-Climbing} displays the results of finger-level matching on the dataset using synthetic PatternMasterPrints. By applying a hill-climbing method on the top five ranked synthetic PatternMasterPrints, they are regenerated from the training set. If the attack success rate remains unchanged within a predefined number of iterations, it indicates that the performance of the synthetic PatternMasterPrints is stabilizing. The hill-climbing process is terminated if the success rate does not improve within 100 iterations. From Figure~\ref{12-Hill-Climbing}, it can be observed that after 300 iterations, the improvement in the success rate of the iteratively generated PatternMasterPrint becomes smaller. Therefore, we set the maximum iteration count, denoted as ${i}_{max}$, to 500.


\section{Evaluation}
In this section, we present the experimental results based on our PrintListener prototype.

\subsection{Experimental Settings}

\textbf{Data Collection.} After obtaining IRB approval, we recruited 65 participants (24 females and 41 males, aged 18 to 30) to participate in our experiments. \red1{Some participants exhibit mild perspiration on their hands as we do not require participants' fingers to be completely dry.} Thus, our \red1{datasets match} the finger characteristics of a larger and more diverse population. Before data collection, the participants were provided with a detailed explanation of the study's purpose and signed an informed consent form. We observed the fingerprint patterns of all ten fingers of the participants and selected 1-6 fingers for each participant to collect fingerprint friction sounds and corresponding fingerprint images. \red1{As imbalanced datasets could introduce bias in the feature weighting for the three fingerprint patterns and lead to skewed predictive results, we chose a different number of fingers for different participants to balance the ratio of the whorl, left loop, and right loop fingerprints at 1:1:1.} We collect friction sounds of 180 fingers (60 each of left loop, right loop, and whorl patterns) and use the ZKLive20R optical fingerprint scanner to get corresponding optical fingerprint images. For each chosen finger, the participant swiped on the screen 25 times \red1{based on their regular usage habits}. The details of compiled datasets are explained in APPENDIX B.

\textbf{Default Setting.} 
The sampling rates of instant messaging software are usually set \red1{from} 16 kHz \red1{to} 44.1 kHz. For consistency and comparability, we set the basic sampling rate at 44.1kHz. Our experiments are conducted in three different environments: a conference room (quiet), an office (slightly noisy), and a playground (noisy). We collected the finger-swiping friction sound on three commercial smartphones: the Google Pixel 4, iPhone 13, and Samsung A20S. By default, we used Google Pixel 4 as the evaluation platform in a conference room.
All smartphones were coated with matte screen protectors. We utilized the open-source NBIS fingerprint matcher ~\cite{NIST} to conduct exhaustive matching of fingerprint images in the FingerPassDB7, Livdet2011 ItalData, and FingerPattern datasets. 

\begin{table*} [!htb]
  \renewcommand{\arraystretch}{1.1}
  \caption{Results by using different classifiers (P: Precision, R: Recall).}
  \centering
  \label{tab: wide_module comparision}
  \normalsize
  \begin{tabular}{lcccc}
    \hline
   \makecell[c]{Module} & Accuracy  & P (left loop/right loop/whorl) & R (left loop/right loop/whorl) &  $F_1$ score\\
    \hline
    YAMnet+KNN  & 0.709 & 0.760   /   0.712   /   0.657 & 0.747   /   0.694  
 /   0.685 & 0.709 \\
    YAMnet+Decision Tree & 0.820 & 0.828 / 0.758 / 0.881 & 0.840 / 0.804 / 0.816 & 0.821\\
    YAMnet+Ramdom Forest   & 0.731 & 0.718 / 0.767 / 0.705 & 0.614 / 0.825 / 0.755 & 0.731 \\
    YAMnet+Adaboost   & 0.776 & 0.778 / 0.751 / 0.797 & 0.762 / 0.715 / 0.851 & 0.775 \\
    \hdashline
    \textbf{VGGish-like+KNN}  & \textbf{0.884} & \textbf{0.939} / \textbf{0.865} / \textbf{0.857} & \textbf{0.915} / \textbf{0.895} / \textbf{0.887} &  \textbf{0.886} \\
    VGGish-like+Decision Tree & 0.766 & 0.777 / 0.791 / 0.736 & 0.800 / 0.672 / 0.825 & 0.767\\ 
    VGGish-like+Ramdom Forest   & 0.739 & 0.711 / 0.779 / 0.725 & 0.752 / 0.835 / 0.632 & 0.739 \\
    VGGish-like+Adaboost   & 0.774 & 0.831 / 0.737 / 0.764 & 0.696 / 0.736 / 0.889 & 0.776\\
    \hdashline
    Resnet34+KNN  & 0.746 & 0.739 / 0.706 / 0.815 & 0.763 / 0.841 / 0.633 & 0.750 \\
    Resnet34+Decision Tree & 0.686 & 0.746 / 0.673 / 0.652 & 0.644 / 0.697 / 0.718 & 0.688\\
    Resnet34+Ramdom Forest   & 0.753 & 0.795 / 0.735 / 0.735 & 0.712 / 0.681 / 0.865 & 0.754\\
    Resnet34+Adaboost   & 0.686 & 0.647 / 0.900 / 0.658 & 0.771 / 0.612 / 0.675 & 0.710 \\ 

    \hline
\end{tabular}
\end{table*}

\begin{figure*}[!t]
\centering
\includegraphics[width=0.92\textwidth]{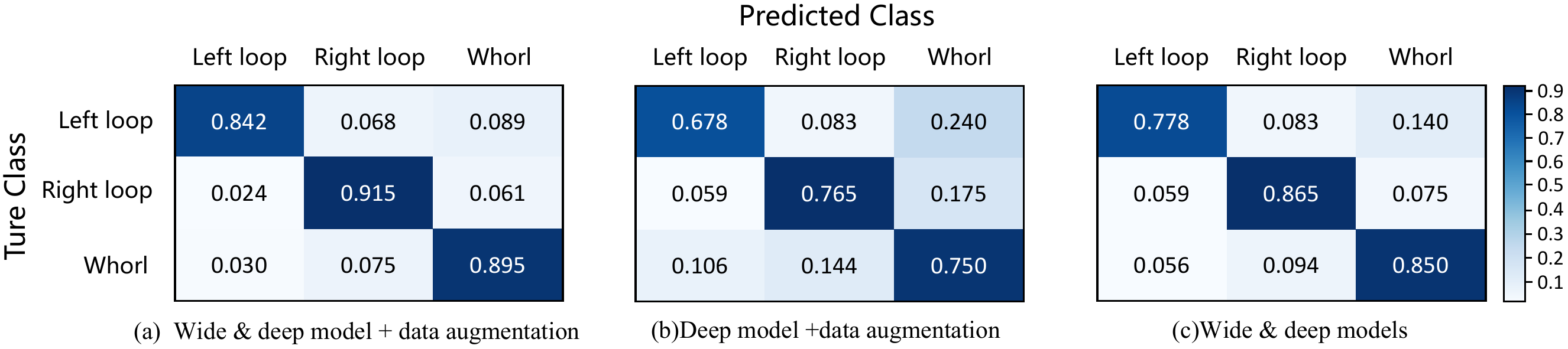}

\caption{\label{fig: confusion matrix} Confusion matrix of different models with/without data augmentation. }
\end{figure*}

\subsection{Metrics}

We employ accuracy, weighted-average precision ($wP$), weighted-average recall ($wR$), and $F_1$ score as evaluation metrics to assess the performance of our fingerprint pattern prediction model. Accuracy measures the proportion of correctly classified samples among all samples. $TP$ (True Positive) is the number of positive samples correctly predicted as positive, $FP$ (False Positive) is the number of negative samples incorrectly predicted as positive, and $FN$ (False Negative) is the number of positive samples incorrectly predicted as negative. For each fingerprint type, we calculate its precision ($TP/(TP+FP)$) and recall ($TP/(TP+FN)$) separately. The precision and recall of the left loop, right loop, and whorl are weighted according to the dataset proportions to obtain $wP$ and $wR$. To comprehensively evaluate the precision and recall performance of our friction sound classification model, we also select $F_1$ score ($2\cdot wP\cdot wR/(wP+wR)$) to balance the two metrics.

The False Acceptance Rate ($FAR$) is the proportion of unauthorized user data mistakenly accepted as genuine users. A lower $FAR$ value signifies increased security, albeit with diminished usability for the authentication system. To evaluate the attack capability of PrintListener within various security settings, we establish three distinct security gradients for the fingerprint matcher ($FAR=1\%, FAR=0.1\%, FAR=0.01\%$). We utilize the open-source NBIS fingerprint matcher ~\cite{NIST} to conduct exhaustive matching of fingerprint images in 
\emph{Dataset-4\_FingerPassDB7}, \emph{Dataset-5\_Livedet2011 ItalData}, and \emph{Dataset-6\_PatternFinger}. Subsequently, we determine the corresponding threshold values for the three $FAR$ settings. 
The ultimate threshold values for the three fingerprint datasets under different $FAR$ settings are presented in Table~\ref{tab: Threshold and TAR}.

We use a weighted attack success rate ($wASR$) to evaluate the effectiveness of PatternMasterPrint. For the $i$-th type of fingerprints, the $wASR$ of generated PatternMasterPrint ${wASR}_{i}$ is calculated as:
\begin{equation}
{wASR}_{i}=\sum_{j=1}^{3} R_{i j}\cdot ASR_{j i},
\label{con:equa9}
\end{equation}
where $i$ and $j$ take the values 1, 2, or 3, representing the left-loop, right-loop, and whorl patterns, respectively. $R_{ij}$ represents the probability of predicting a fingerprint with $i$-th type features as the $j$-th class fingerprint. $ASR_{j i}$ denotes the attack success rate of the $j$-th PatternMasterPrint against the $i$-th type of fingerprint.

\subsection{Fingerprint Pattern Prediction}
In this section, we compare the performance of different wide \& deep modules on friction sound classification and then select the best one (our wav2pattern model) for the subsequent evaluation.
After determining the chosen combined network, we assess the effects of data augmentation, sampling rate, environmental noise, and friction sound collection equipment. 

\subsubsection{Impact of Combined Network}
We evaluate the impact of various joint models on \emph{Dataset-1}. We compare our VGGish-like model with a sound-based convolutional neural network YAMnet~\cite{yamnet}, and an image-based CNN model ResNet34 \cite{he2016deep}. By utilizing combination and permutation, we can obtain 12 distinct wide and deep joint classification networks. The results of our experiments are presented in Table~\ref{tab: wide_module comparision}, which displays the classification accuracy, precision, recall, and F1 scores of each joint network. The results show that our wav2pattern network outperforms the others with an accuracy of 88.4\%. Moreover, the classifier does not show a significant bias towards any particular type of fingerprint.
Figure~\ref{fig: confusion matrix}a shows the confusion matrix of wav2pattern. We compared the classification performance without the wide module (KNN) as shown in Figure~\ref{fig: confusion matrix}b. When predicting without the joint use of the wide module, the classification accuracy is 73.09\%, which decreased by approximately 15\%. Since our wav2pattern combined network achieves the best performance, we use wav2pattern for the subsequent evaluation.

\begin{figure*}[!t]

\center{\includegraphics[width=0.6\textwidth]{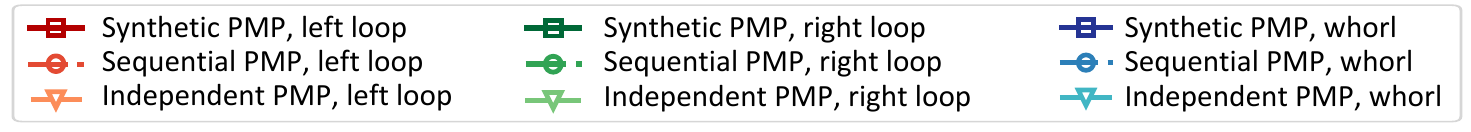}}

\begin{minipage}[t]{0.31\textwidth}
\centering
\subfloat[][PatternFinger]{\label{3-Fingerpattern-FAR01}\includegraphics[width=1\textwidth]{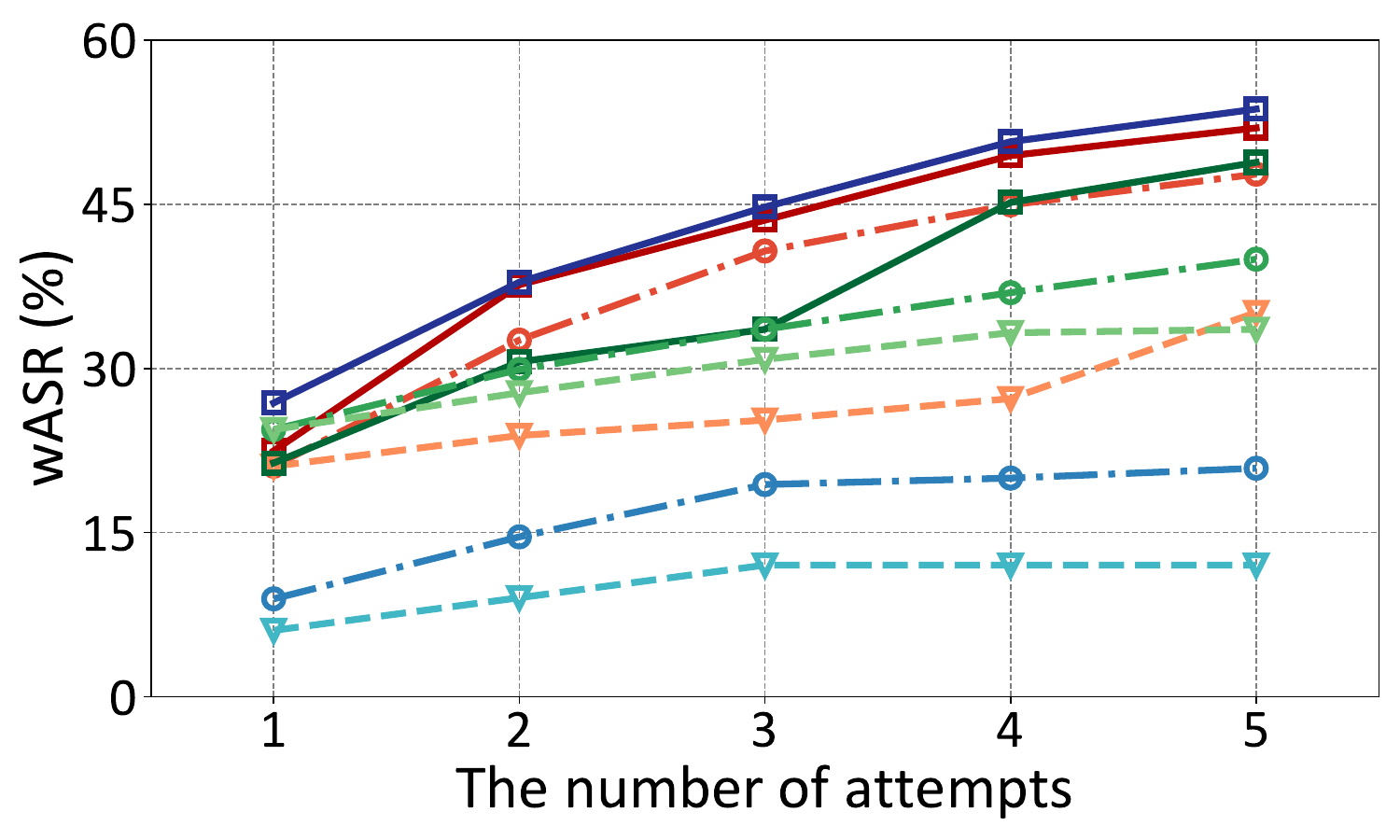}}
\end{minipage}
\hfill
\begin{minipage}[t]{0.31\textwidth}
\centering
\subfloat[][FingerPassDB7]{\label{2-Fingerpass-FAR01}\includegraphics[width=1\textwidth]{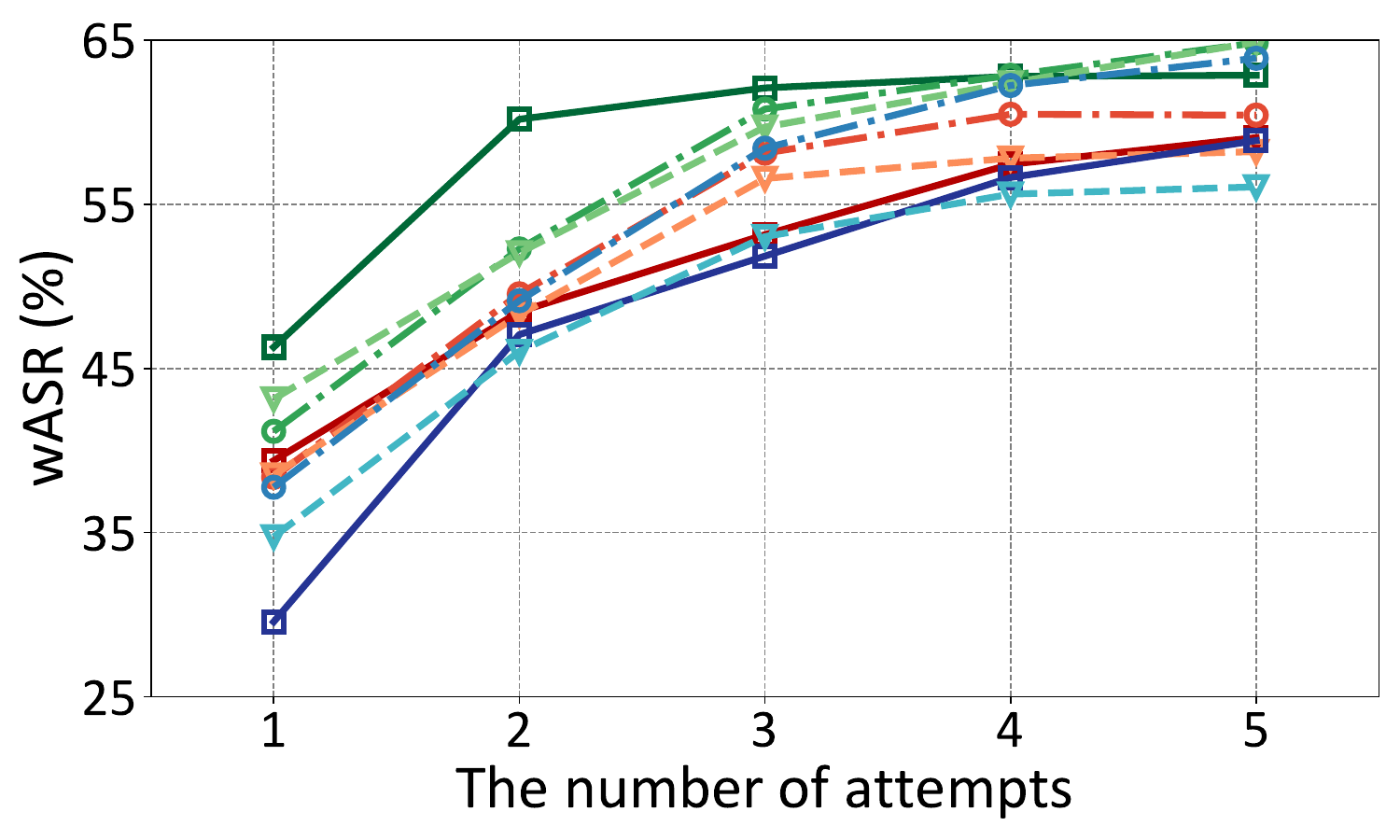}}
\end{minipage}
\hfill
\begin{minipage}[t]{0.31\textwidth}
\centering
\subfloat[][Livedet2011 ItalData]{\label{1-livdet-FAR01}\includegraphics[width=1\textwidth]{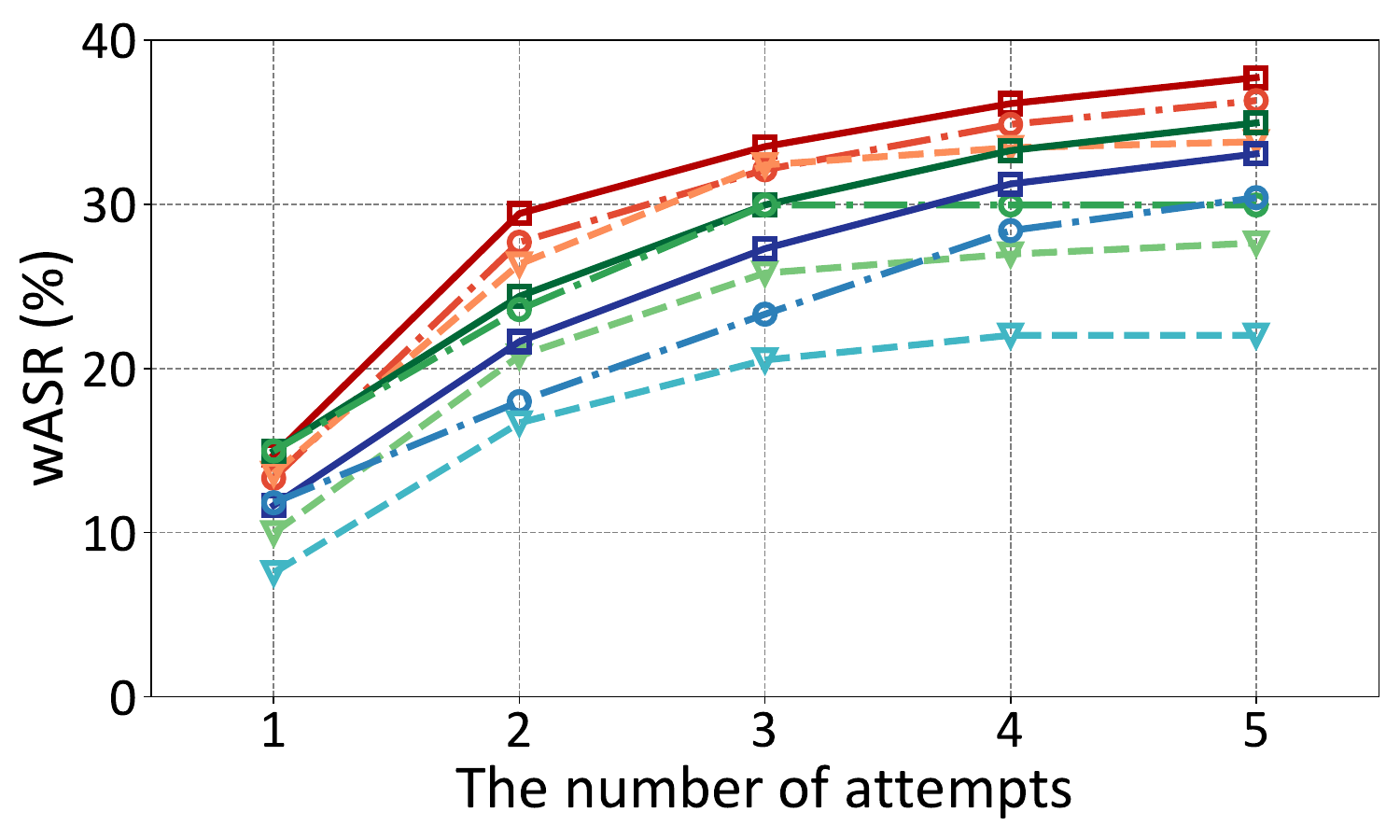}}
\end{minipage}

\caption{The $wASR$ of three methods for generating PatternMasterPrint on different datasets with the $FAR$ set at 0.1\%.}
\label{fig: FAR_on_3datasets}
\end{figure*}

\subsubsection{Impact of Data Augmentation}
To evaluate the impact of data augmentation on PrintListener, we compare the accuracy of wav2pattern on \emph{Dataset-1} before data augmentation (a total of 9,000 friction segments), as shown in Figure~\ref{fig: confusion matrix}a and ~\ref{fig: confusion matrix}c. Wav2pattern achieves an average classification accuracy of 83.09\% without data augmentation, while data augmentation increased the accuracy by 5\%. We attribute this improvement to the variation in users' sliding speed and pressure. 
\red1{Our data augmentation technique, which involves applying pitch shifting and time stretching, helps to alleviate the impact of sliding speed and pressure on the recorded sounds, thereby improving the accuracy of our results.}

\subsubsection{Impact of Sampling Rate}
\begin{figure}[!t]   
\center{\includegraphics[width=7.5cm] {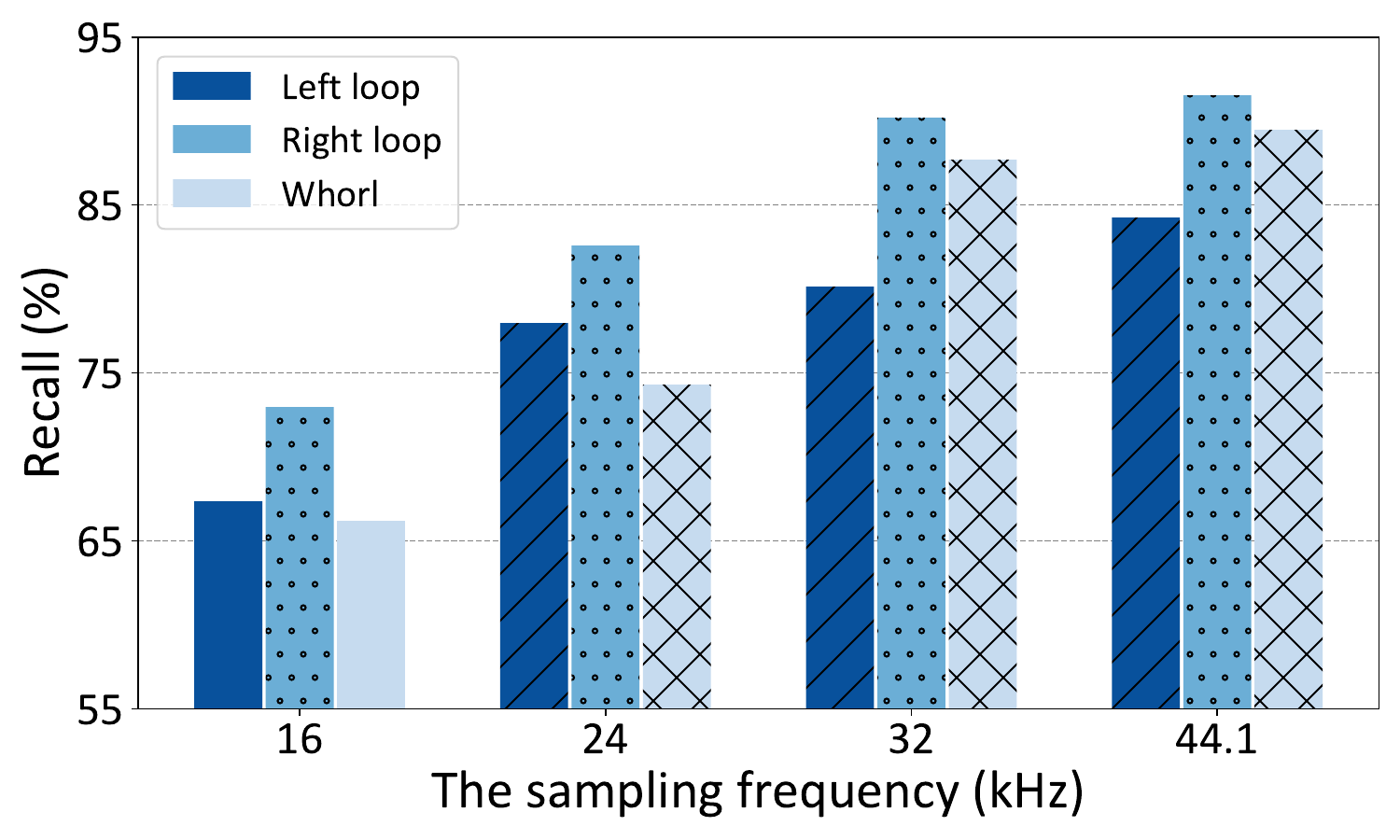}}   
\caption{\label{fig: Auc_p} Impact of sampling frequencies.}   
\end{figure}

\red1{We evaluate the performance of PrintListener under four common sampling rates used in current audio and video social networking software}. The friction segments at different sampling rates are obtained by downsampling 45,000 data points from \emph{Dataset-1}. As shown in Figure~\ref{fig: Auc_p}, the recall in classifying the fingerprint pattern gradually decreases as the sampling rate decreases. This is because the effective information contained in the audio signal inevitably reduces as the cutoff frequency decreases. Specifically, high-frequency sound segments contain more detailed information than low-frequency sound segments. Notably, the classification accuracy of sound at 44.1 kHz and 32 kHz sampling rates does not exhibit a significant distinction. 32 kHz is a commonly used sampling rate in most audio and video social networking software.

\subsubsection{Impact of Environments}
We conduct a comparison of two noise environments:  (a laboratory, \emph{Dataset-2A}) and a corridor (\emph{Dataset-2B}) with a quiet environment (an office, \emph{Dataset-1}).
As shown in Table~\ref{tab: environment comparison}, the weighted-average precision ($wP$), weighted-average recall ($wR$), and $F_1$ score slightly decreased when friction sounds are affected by low noise. Even in noisy corridors with speech or crowd-walking sounds, the $wP$ is 87.5\%, a decrease of only 1.2\%.
This indicates that our model is insensitive to low ambient noise since the user's fingerprint features are widely distributed in different frequency bands.

\subsubsection{Generalization}
We conduct generalization experiments to evaluate the robustness of wav2pattern to hardware fingerprint and quality. We select 4,500 friction sound segments (\emph{Dataset-3}) collected from a Samsung A20s phone with \red1{a long-term used microphone}. Then, we examine whether the joint network trained on Pixel 4 and iPhone 13 can accurately classify them. The result demonstrates a certain degree of robustness to microphone hardware fingerprint and quality, achieving an accuracy of 80.32\%. The accuracy on the Samsung A20s only decreases by approximately 8\% compared to that on Pixel 4 and iPhone 13, indicating that hardware fingerprint and quality slightly impact the friction sound features.

\begin{figure*}[!t]   
\center{\includegraphics[width=18.5cm] {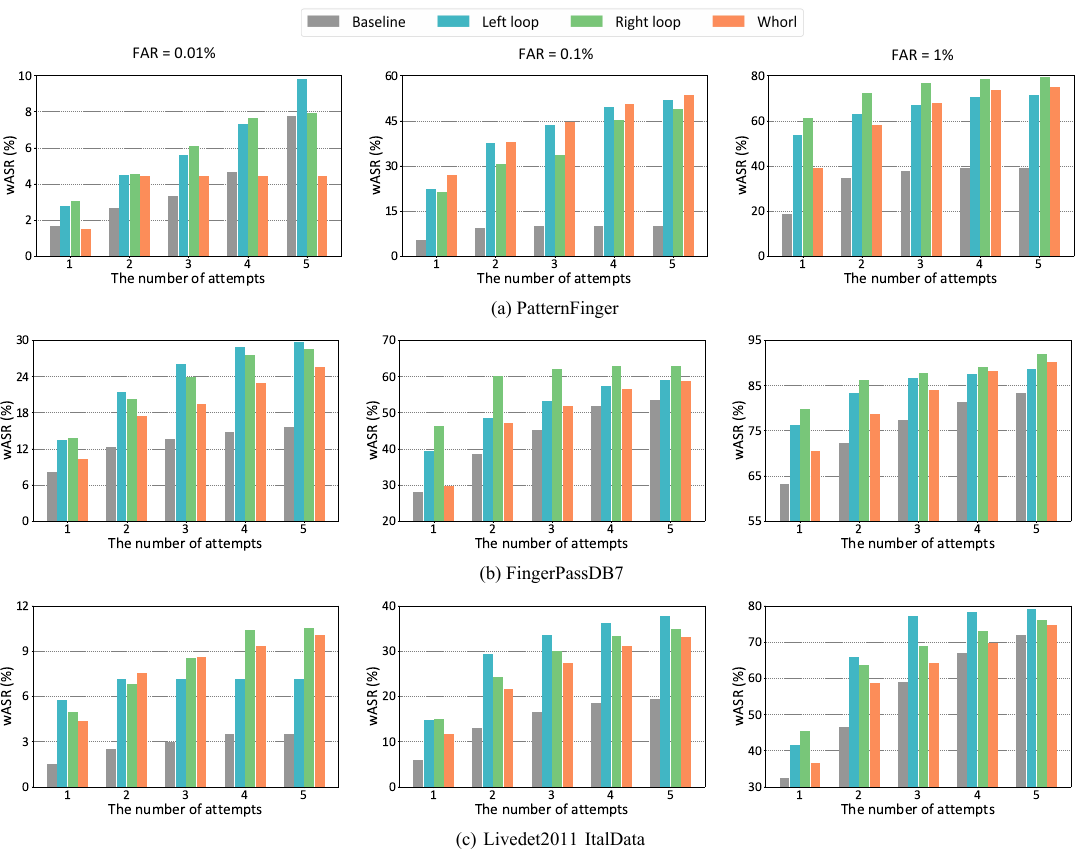}}   
\caption{\label{baseline_compare} The $wASR$ of synthetic PatternMasterPrint and baseline on different datasets.}   
\end{figure*}

\subsection{PatternMasterprint Attack}

\subsubsection{Impact of Fingerprint integrity}
Figure~\ref{fig: FAR_on_3datasets} illustrates the weighted attack success rates ($wASR$) of three methods for generating PatternMasterPrint on partial fingerprints (\emph{Dataset-5\_FingerPassDB7}) and 
complete fingerprints (\emph{Dataset-4\_PatternFinger} and \emph{Dataset-6\_Livedet2011 ItalData} ), with the $FAR$ set at 0.1\%. 
It can be observed that the $wASR$ of partial fingerprints is significantly higher than that of complete fingerprints. This is because partial fingerprints are more prone to erroneous matching due to the loss of information entropy. Moreover, the $wASR$ of synthetic PatternMasterPrints is generally higher than that of sequential PatternMasterPrints and independent PatternMasterPrints. 
Therefore, we use synthetic PatternMasterPrints as an example to discuss the impact factors of the attack success rate of PrintListener.


\begin{table} [!t]
\renewcommand{\arraystretch}{1.1}
  \caption{\red1{Impact of environments.}}
  \centering
  \label{tab: environment comparison}
  \normalsize
  \begin{tabular}{cccc}
    \hline
   Environment & $wP$ & $wR$  & $F_1$ score\\
    \hline
    Office  & 0.887 & 0.884 & 0.886\\    
    Laboratory & 0.880 & 0.878 & 0.879\\    
    Corridor   & 0.875 & 0.874 & 0.874 \\
    \hline
\end{tabular}
\end{table}

\subsubsection{Impact of FAR security setting}
The impact of the $FAR$ value on $wASR$ of PatternMasterPrint attacks in \emph{Dataset-4\_PatternFinger} is demonstrated in Figure~\ref{baseline_compare}a. 
We can see that the $wASR$ decreases at a lower $FAR$ value setting (higher security), while more test subjects can be successfully attacked at a higher $FAR$ value setting (lower security). 
When the $FAR$ is set to 0.1\% (a value that balances security and usability), PrintListener can successfully attack 52\% of users with the left loop fingerprints, 48.8\% of users with the right loop fingerprints, and 53.7\% of users with the whorl fingerprints within 5 attempts.  
When the $FAR$ is set to 0.01\%, the $wASR$ is 9.8\% for users with the left loop fingerprints,  7.9\% for users with the right loop fingerprints, and 7.8\% for users with the whorl fingerprints within 5 attempts. Therefore, PrintListener will indeed pose a huge threat to real fingerprint authentication systems.




\subsection{Baseline Comparisons}
Finally, we compare the performance of PrintListener and the related fingerprint attack method, \ie, MasterPrint~\cite{roy2017masterprint}. 
Due to the lack of open sourcing of DeepMasterPrint~\cite{bontrager2018deepmasterprints} and the unavailability of critical information regarding its training process, it is inaccessible to set DeepMasterPrint as a baseline for comparison with PrintListener. Besides, DeepMasterPrint only conducted attack experiments on partial fingerprint images, not applicable to this work.


\textbf{Partial Fingerprints.}
Firstly, we compare the $wASR$ of Printlistener and MasterPrint for attacking \emph{Dataset-5\_FingerPassDB7} at three $FAR$ settings. 
As shown in Figure~\ref{baseline_compare}b,  the MasterPrint sequences selected through pattern prediction generally have higher attack success rates than those without pattern prediction. At a $FAR$ of 0.1\%, Printlistener demonstrates an average improvement of 37.0\% in $wASR$ compared to the attack success rate of Masterprint in only one attempt. At the highest security $FAR$ setting of 0.01\%, PrintListener achieves the average $wASR$ of 27.9\% within 5 attempts which is 1.8 times the attack success rate of Masterprint.

\begin{figure*}[!t]   
\center{\includegraphics[width=18cm]  {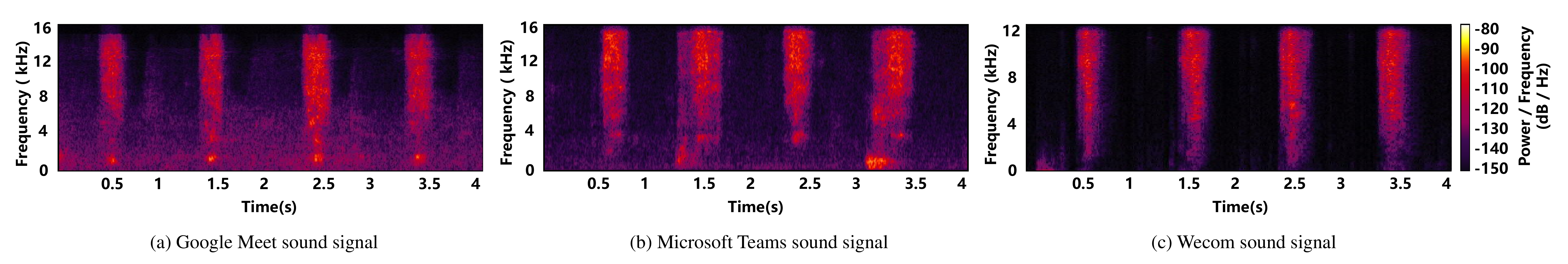}}   
\caption{\label{diff app} Frictional sound on different apps.}   
\end{figure*}

\textbf{Complete Fingerprints.} 
Next, we compare the $wASR$ of Printlistener and MasterPrint for attacking \emph{Dataset-6\_Livedet2011 ItalData} at three $FAR$ settings. As shown in Figure~\ref{baseline_compare}c, when the $FAR$ is 0.1\%, Printlistener can successfully attack 37.7\% of left loop fingerprints, 34.9\% of right loop fingerprints, and 33.0\% of whorl fingerprints within 5 attempts, representing a significant improvement of 93.3\%, 79.0\%, and 69.2\%, respectively, compared to the performance of MasterPrint.
At the highest security $FAR$ setting of 0.01\%, 
PrintListener achieves an average $wASR$ of 9.3\% within 5 attempts which is 2.6 times the attack success rate of Masterprint.



\section{Discussion}

\begin{table} [!t]
  \caption{Sampling rate statistics for Social networking applications.}
  \centering
  \label{tab: sample_rate_of_apps}
  \normalsize
  \renewcommand\arraystretch{1.1}
  \begin{tabular}{c|c}
    \hline
      Apps & Sampling rates (kHz)\\    
    \hline
      Skype & 8 / 12 / 16 / 24\\    
      FaceTime & 8 / 12 / 16 / 24\\     
      Google Meet & 24 / 32  \\      
      Microsoft Teams& 16 / 32\\
      Wecom  & 16 / 24 \\
    \hline 
    
  \end{tabular}
\end{table}

\subsection{Attack Feasibility via Social Networking Apps}

\red1{The prevalent sampling rates used by popular audio and video social networking Apps are shown in table \ref{tab: sample_rate_of_apps}. It can be observed that the majority of software applications maintain sampling rates above 16 kHz, which is the minimum sampling rate in our evaluation. As stated in the evaluation, when the sampling rate is sufficiently high, audio compression does not interfere with the extraction of friction sound signal features. When an attacker covertly records the transmitted audio signal in the background, the spectrogram retains distinct features that can be utilized. This indicates that our inference of friction sound features can be extended to other audio and video software applications. We collect swipe sounds of users on several social software. The spectrograms are shown in Figure~\ref{diff app}.}

\subsection{Defense}
A simple countermeasure to prevent the leakage of finger friction sound containing fingerprint features is to correct some users' habits. For example, users try not to swipe their fingers on the phone screen when making audio and video calls on social media platforms. However, it is difficult to avoid not performing the swiping operation in some scenarios, \eg, engaging in online gaming on mobile phones or tablets through social applications. In addition, we suggest that audio/video social and communication apps be limited to lower audio sample rates. As shown in Figure \ref{fig: Auc_p}, the recall in classifying the fingerprint pattern gradually decreases as the sampling rate decreases. Limiting the audio sampling rate can reduce the leakage of fingerprint information to a certain extent. \red1{Additionally, audio/video social apps can destroy finger frictional sound features with automatic speech noise reduction, or implement pop-up reminders to caution users to be careful when performing swiping operations while the microphone is in use.}

\begin{figure}[!t]  
\center{\includegraphics[width=8.9cm]  {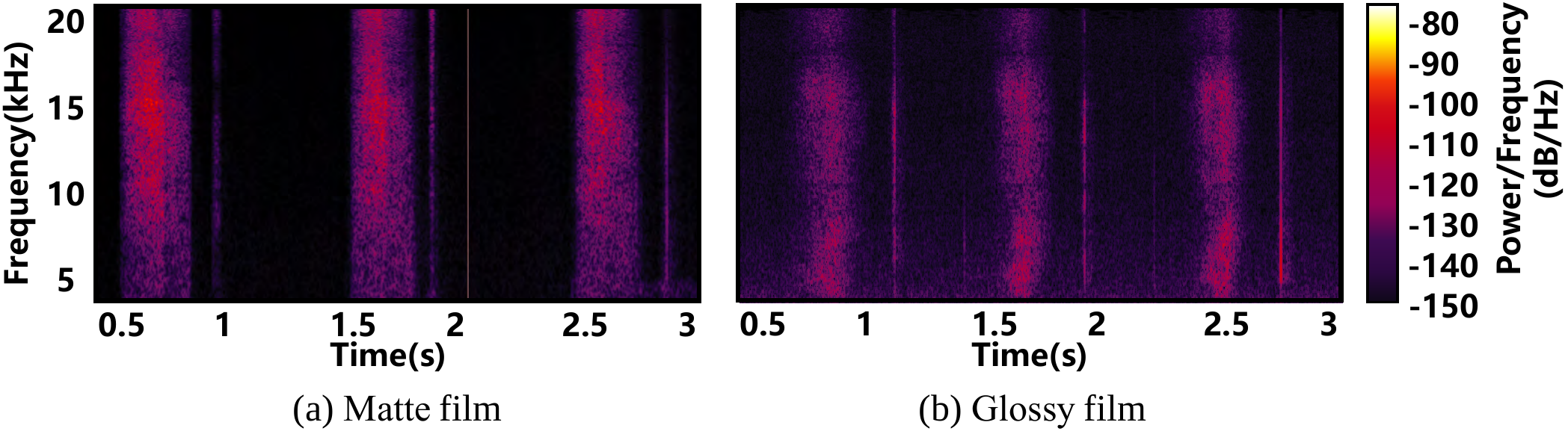}}   
\caption{\label{diff film}  Frictional sound on different films.}   
\end{figure}

\subsection{Limitations}
Different screen protectors on mobile devices may affect the sliding friction sound characteristics, weakening our model's classification accuracy and robustness. Screen protectors can be divided into matte film and glossy film according to the processing technology.
When users interact with mobile devices, their fingers will \red1{come into contact} with the screen surfaces, leaving behind fingerprint marks in the form of grease or sweat stains. These marks can be exploited by attackers to create fake fingerprints based on the imprints \cite{bumbrah2016emerging}. As a solution, matte films have gained popularity recently due to their ability to prevent fingerprint leakage and reduce glare \cite{wang2022heterogeneously, MattevsGlossy, mattefilm}.
Our experiments evaluated the finger friction sound produced on the matte film. At the same time, we also tested the friction effects on the glossy film, the spectral energy of frictional sound is a little weak (Figure~\ref{diff film}). However, with longer use, the surface of the glossy film will become rough, and the friction sound will be more pronounced.


\section{Related Work}
\subsection{Fingerprint Template Attack}
An Automated fingerprint identification system (AFIS) recognizes individuals by comparing the spatial distribution of ridge structures and detail points~\cite{cao2018automated, zhang2021multi}. Although AFIS is considered to be efficient and practical, they are also vulnerable to presentation attack. This attack usually uses some commonly available materials (\eg, gelatin, wax, liquid latex, \etc) to create fake synthetic fingerprints, such as 2D fingerprint images and 3D fake fingers. Cao \etal~\cite{cao2016hacking} successfully spoofed the Samsung Galaxy S6 and Huawei Honor 7 using 2D fingerprint images printed on special paper. The research of CISCO~\cite{3dfinger} chose fabric glue to create 3D fake fingers, achieving an average success rate of 80\% in bypassing AFIS. However, in practical scenarios, collecting one clear fingerprint image of the victim is challenging. 
To avoid this problem, Roy \etal~\cite{roy2017masterprint} generated Masterprint based on a large set of fingerprint data that could match one or more fingerprint templates of a vast number of users by chance. 
Bontrager \etal ~\cite{bontrager2018deepmasterprints} leveraged a GAN network to generate image-level MasterPrint, called DeepMasterPrint. However, the attack success rate of Masterprint and DeepMasterPrint is very low at a high-security level.
Chen \etal~\cite{cheninfinitygauntlet} discovered a design flaw in the smartphone fingerprint authentication system that allows unlimited attempts at fingerprint authentication, thereby enabling brute-force attacks without requiring any prior knowledge of the victim. 
Some simple measures can make up for this design flaw, \eg,  checking whether a cancellation happens during each fingerprint authentication.

\subsection{User Information Inferring via Acoustic}
The acoustic signals related to the user contain some important user information. There have been some studies on stealing users' private information through acoustic signals.
Keyboard recognition based on acoustic emanation has been studied in~\cite{asonov2004keyboard,berger2006dictionary,zhuang2009keyboard,zhu2014context,wang2014ubiquitous,liu2015snooping}. These approaches utilized slightly different keystroke sounds between different keys, or the difference in arrival time of the keystroke sounds of different keys to infer the user's keystroke information. Among them, ~\cite{zhu2014context,wang2014ubiquitous,liu2015snooping}
leveraged a malicious APP on a mobile device to identify keystrokes on a nearby keyboard, thereby eavesdropping on the user's keyboard input. Arp \etal ~\cite{arpprivacy} explored the functionality, prevalence, and technical limitations of ultrasonic beacons embedded in audio and tracked users using mobile device microphones. PatternListener~\cite{zhou2018patternlistener} and UltraPIN~\cite{liu2021ultrapin} utilized the user's smartphone to emit ultrasonic signals to track the user's finger movement and infer the unlock pattern and PIN respectively. PrintListener is the first work to infer users' fingerprint information using acoustic signals.

\section{Conclusion}
In this paper, we uncover a new side-channel attack on fingerprint authentication and propose PrintListener, which utilizes users' fingertip-swiping actions on the screen to extract fingerprint features and synthesizes powerful PatternMasterPrint sequences for fingerprint dictionary attacks. Extensive experimental results in real-world scenarios demonstrate that Printlistener can attack up to 26.5\% of partial fingerprints and 9.3\% of complete fingerprints within five attempts at the highest security FAR setting of 0.01\%, which far exceeds the attack potency of MasterPrint.

\section*{Acknowledgments}
We would like to thank the anonymous reviewers for their helpful feedback. Man Zhou's work was partially supported by the NSFC under Grants 62202180. Qian Wang's work was partially supported by the NSFC under Grants U20B2049 and U21B2018. Qi Li's work was partially supported by the NSFC under Grant 62132011. Xiaojing Ma's work was partially supported by the NSFC under Grants 62272175.

\bibliographystyle{IEEEtran}
\bibliography{PrintListener2}


\appendix

\begin{table} [htb]
\renewcommand{\arraystretch}{1.1}
  \caption{The parameters of our VGGish-based CNN model.}
  \centering
  \label{tab:deepmodel}
  \normalsize
  \begin{tabular}{cccc}
    \hline
     Layer Type & Kernel & Output Shape  \\
    \hline   
     Conv1 + BN +Relu1+ Pool1  & 64 & 48×32×64 \\      
     Conv2 + BN +Relu2 + Pool2  & 128 & 24×16×128 \\      
     Conv3\_1 +Relu3  & 256 & 24×16×256 \\
     Conv3\_2 +Relu4 + Pool3  & 256 & 12×8×256 \\
     Conv4\_1 + Relu5  & 256 & 12×8×512 \\
     Conv4\_2 + Relu6 + Pool4  & 256 & 16×4×512 \\          
     Flatten Layer  & / & 12288  \\
     Inner product (FC1+FC2+FC3)  & / & 128  \\
    \hline
\end{tabular}
\end{table}

\begin{figure*}[!t]   
\center{\includegraphics[width=0.7\textwidth] {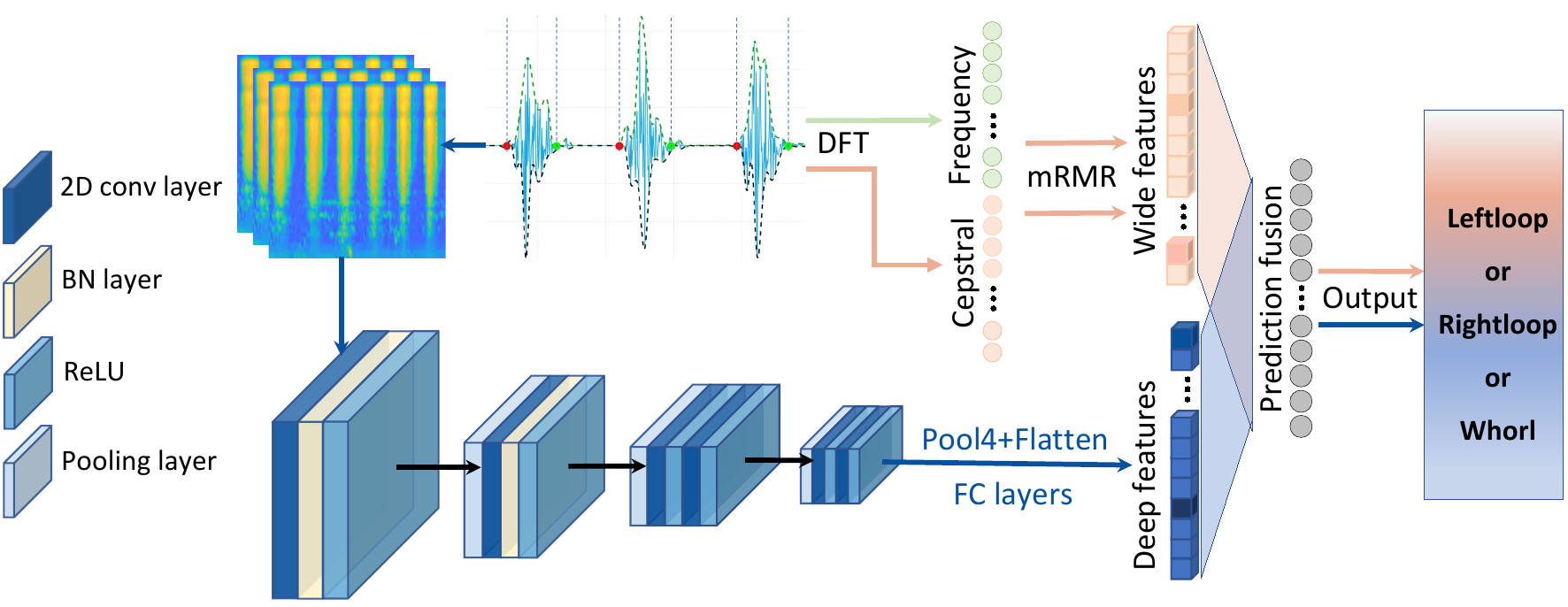}}   
\caption{\label{fig:VGGish} The architecture of our \red1{wav2pattern} model.}   
\end{figure*}

\subsection{Implementation Details of VGGish-based CNN Encoder}
Based on data extracted from 120 finger friction sound segments collected from 65 users, we train our model with a total of 10,800 data points. 
We utilize a computer with the GPU of Tesla T4 and TensorFlow as the backend. The network architecture and parameters are shown in Figure ~\ref{fig:VGGish} and Table~\ref{tab:deepmodel}.
The network optimization is performed using mini-batch gradient descent (MBGD), with a fixed learning rate of 0.001 and weight decay parameter of 0.0005. The batch size is set to 128, and batch normalization and rectified linear activation functions are applied after each convolutional layer. The model is trained for 1000 iterations. Although the VGGish model is trained on a limited dataset, experimental results demonstrate that it could effectively transfer the discriminative knowledge of audio to finger friction sound for classification modeling. Additionally, we have considered the influence of different environments and recorded the audio of users sliding their mobile phone screens in a noisy office environment.

\RestyleAlgo{ruled}
\SetKwComment{Comment}{/* }{ */}
\begin{algorithm}[!t]
	\caption{Random Restart Hill-Climbing.} 
        \label{alg:Hill-Climbing}
	\SetKwData{Left}{left}\SetKwData{This}{this}\SetKwData{Up}{up}
	\SetKwFunction{Union}{Union}\SetKwFunction{FindCompress}{FindCompress}
	\SetKwInOut{Input}{input}\SetKwInOut{Output}{output}
	
	\Input{$sPMP = ({sPMP}_{1},{sPMP}_{2},{sPMP}_{3},{sPMP}_{4},{sPMP}_{5})$; $Fp{}$; $outFp{}$; ${j}_{max}$; ${S}_{1}, {S}_{2}, {S}_{3}, {S}_{4}, {S}_{5}$ ; 
		${Temp}_{S}$ \Comment*{Temporary storage template}
		
	}
	\Output{Top 5 Synthetic PMPs}
	
	\For{$i=1$ to $5$}{
		Calculate ${ASR}_{max}$ and ${smp}_{i}$\;
	$j \gets 0$\;
		\While{$j \leq {j}_{max}$}{
	    $j$++\;
	    a) Modify a random detail to ${Temp}_{S}$\;
	    b) Add a randomly generated detail to ${Temp}_{S}$\;
	    c) Randomly replace details in existing detail sets\;
	    d) Randomly delete a detail from an existing detail set\;
	    Calculate $temp_{asr} = ASR(temp_S)$\;
		\If{$temp_{asr} \geq bestasr_i$}{
			$S_i = temp_S$\;
			$bestasr_i = temp_{asr}$\;
		    }
	
	    }
		\If{$i$ == 5}{
			Break\;
		}
	}
\end{algorithm}

\subsection{Datasets}\label{Datasets}
The following datasets are used in our experiments, where datasets 1-5 are produced from our data collection. Datasets 6 and 7 are typically open-source datasets as the baseline.

\emph{Dataset-1}: We select Google Pixel 4 and iPhone 13 as the devices for finger friction sound collection. 
Sixty-five users (a total of 180 fingers) slide their fingers on each phone screen 25 times. i) \emph{Dataset-1A}: In this dataset, the friction sounds of 180 fingers sliding on Pixel 4's screen are collected. After undergoing audio denoising, segmentation, and data augmentation, a total of $180 \times25 \times5=22,500$ finger friction sound segments are obtained. ii) \emph{Dataset-1B}: To increase the robustness of the classification model, we also collect Dataset-1B consisting of 22500 friction sound segments using iPhone 13.
Due to microphone fingerprints, the sound collected by different microphones may have slight differences. To address this potential issue, we collect data using two different devices in a quiet office, resulting in a total of $22,500 \times 2=45,500$ friction sound segments.

\emph{Dataset-2}: To evaluate the impact of noise, we additionally record the friction sound of 90 fingers on a Google Pixel 4 in two different noise environments. The first environment is a laboratory with mild noise, including the sounds of typing and a printer, and the second environment 2 is a moderately noisy corridor, including sounds of walking and light conversation. These environments corresponded to \emph{dataset-2A} and \emph{dataset-2B}, respectively. After data pre-processing, dataset-2 consists of a total of $90 \times2 \times25 \times5=22,500$. 

\emph{Dataset-3}:
To evaluate the generalization performance of our model on friction sounds collected from different devices and microphones, we collect a  set of friction sounds using the Samsung A20s. Specifically, we aimed to test the robustness of our model to untrained device microphone fingerprints. 180 fingers were slid 25 times on the Samsung A20s, resulting in dataset-3 of 4500 ($180 \times25$) friction sound segments. As dataset-3 was only used for generalization testing, we did not perform any data augmentation on these data points. 

\emph{Dataset-4\_PatternFinger}: This is the complete fingerprint dataset, where the data is collected by an optical sensor. There are 2160 complete fingerprint images from 180 fingers with 12 complete fingerprint impressions per finger. Each complete fingerprint image is $400 \times 300$ pixels in size and 500 dpi in resolution. To evaluate the impact of fingerprint patterns on the attack success rate, \emph{dataset-6} is further subdivided into \emph{dataset-4A}, \emph{dataset-4B}, and \emph{dataset-4C}, corresponding to whorl, left loop, and right loop patterns, respectively. Each subset includes 720 ($60\times12$) fingerprint images from 60 fingers.

\emph{Dataset-5\_FingerPassDB7}: In order to evaluate the impact of fingerprint image integrity, we selected a subset of the fingerPassDB7 fingerprint dataset. This is a  natural partial fingerprint dataset with data collected by the Authentic AES3400 capacitive sensor. There are 8640 partial fingerprints  from 720 fingers. Each fingerprint image is $144\times144$ pixels and 500 dpi in resolution. \emph{Dataset-5} can be further subdivided into \emph{dataset-5A}, \emph{dataset-5B}, \emph{dataset-5C}, and \emph{dataset-5D}, corresponding to whorl, left loop, right loop, and arch patterns, respectively.

\emph{Dataset-6\_Livedet2011 ItalData}: This is the complete fingerprint dataset, where the data is collected by an optical sensor. There are 2000 complete fingerprint images from 400 fingers with 5 complete fingerprints per finger, each complete fingerprint image is $680\times480$ pixels in size and has the same resolution of 500 dpi. Based on the whorl, left loop, right loop, and arch patterns, \emph{dataset-6} can be divided into \emph{dataset-5A}, \emph{dataset-5B}, \emph{dataset-5C}, and \emph{dataset-5D}.

\end{document}